\documentclass{aa} 
\usepackage{natbib}
\bibpunct{(}{)}{;}{a}{}{,} 

\usepackage{graphicx}
\usepackage{amsmath}    
\usepackage[colorlinks=true,
            linkcolor=blue,
            citecolor=blue,
            urlcolor=blue]{hyperref}
\usepackage{bm}
\usepackage{newtxtext,newtxmath,siunitx, ulem}

\newcommand{\bqn}{\begin{eqnarray}}
\newcommand{\eqn}{\end{eqnarray}}
\newcommand{\nb}{\nonumber}

\begin{document} 

   \title{Signature of iron line profile from a Kerr-like wormhole}
    \titlerunning{Iron line profile for a Kerr-like wormhole}

   \author{Cheng Liu\inst{1}
          \and
          Hoongwah Siew\inst{2}
          \and
          Hong-Xuan Jiang\inst{1}
          \and
          Yosuke Mizuno\inst{1,2,3}
          \and
          Tao Zhu\inst{4}
          }
   \authorrunning{Liu et al.}

    \institute{Tsung-Dao Lee Institute, Shanghai Jiao Tong University, 1 Lisuo Road, Shanghai 201210, China  \\
               \email{liuc09@sjtu.edu.cn, mizuno@sjtu.edu.cn}
        \and
              School of Physics and Astronomy, Shanghai Jiao Tong University, 800 Dongchuan Road, Shanghai 200240, China 
        \and
              Key Laboratory for Particle Astrophysics and Cosmology (MOE) and Shanghai Key Laboratory for Particle Physics and Cosmology, Shanghai Jiao Tong University, Shanghai 200240, China
        \and
              Institute for Theoretical Physics \& Cosmology, Zhejiang University of Technology, Hangzhou 310023, China \\
              \email{zhut05@zjut.edu.cn}
        }
 
  \abstract 
   {Broad and skewed iron K$\alpha$ emission lines in the X‑ray spectra of accreting black holes encode key information about space-time geometry in the innermost disk regions. While relativistic reflection models that assume the Kerr metric have yielded spin measurements for dozens of black hole candidates, horizonless alternatives, particularly traversable “Kerr‑like” wormholes, can reproduce many observational signatures of black holes and thus challenge the interpretation of these data.
}
   {We aim to develop and apply a new relativistic reflection framework that explicitly incorporates Kerr‑like wormhole geometries to predict the iron line distortions introduced by wormhole throat effects and assess the feasibility of distinguishing event horizons from horizonless throats using current and future X‑ray observations.
}
   {Building on a custom ray‑tracing subroutine for Kerr-like wormhole space-times, we implemented two custom \textsc{XSPEC} modules, \texttt{kwline} for isolated $\delta$‑function line profiles and \texttt{kwconv} for full rest‑frame reflection spectra, parameterized by spin, throat radius, and shape‐function coefficients. We computed a dense grid of line profiles over the wormhole parameter space, embedded them within multicomponent reflection models, and generated synthetic Nuclear Spectroscopic Telescope Array (\textit{NuSTAR}) spectra that incorporate realistic response matrices and backgrounds. By fitting these simulated data with canonical Kerr black hole based models and examining fit quality and residual structure, we quantified the deviations attributable to wormhole geometries.
}
{We find that Kerr-like wormholes systematically produce narrower Fe K$\alpha$ lines with suppressed red wings and angle-dependent morphological shifts as the throat parameter, $\lambda,$ increases. In spectral fitting simulations of a high-flux 50 ks \textit{NuSTAR} exposure (mocked with $\lambda=0.9, a_*=0.998$), standard Kerr black hole convolution models (\texttt{kerrconv}) can effectively mimic the wormhole spectrum with high statistical precision. Conversely, fitting with the self-consistent, angle-dependent reflection model \texttt{relxillCp} results in a formal statistical failure with pronounced structured residuals and unphysical parameter pegging, such as the emissivity index $q_{\rm in}$ reaching its limit of 10. These results indicate that while Kerr-like wormholes and black holes exhibit significant observational degeneracy under simple convolutional modeling, they can be distinguished through more rigid, self-consistent reflection frameworks.
}
{ We conclude that large-throat wormholes are potentially detectable in high-quality X-ray spectra, provided the analysis moves beyond post-processing approximations toward fully consistent relativistic reflection models.
}

\keywords{black hole physics -- gravitation -- relativistic processes -- line profiles -- X-rays binaries}

\maketitle

\section{Introduction} \label{sec:intro}

Accreting compact objects from stellar‑mass black holes in X‑ray binaries to the supermassive engines at the centers of active galactic nuclei (AGNs) efficiently convert gravitational potential energy into hard X-ray radiation via a corona of hot, optically thin plasma enveloping an optically thick accretion disk. A fraction of coronal photons irradiates the disk surface, where they are absorbed by bound electrons, scattered by free electrons, and reemitted through fluorescence and Compton scattering, yielding a rich reflection spectrum: fluorescent emission lines, absorption edges, and the characteristic Compton hump peaking at 20–30~keV \citep{1991A&A...247...25M, Ross:2005dm, Garcia:2013oma}.

The neutral Fe K$\alpha$ line, the most prominent of the emission lines visible in the X-ray spectra of AGNs, comprising K$\alpha_1$ (6.404 keV) and K$\alpha_2$ (6.391 keV) components in a 2:1 intensity ratio due to spin–orbit splitting, dominates the narrow-line flux in the 6–7~keV band {\citep{PhysRevA.56.4554}}. In the rest frame, each component is intrinsically narrow ($\Delta E\lesssim1\,$eV), but in observed spectra, the blend appears broadened and skewed by Doppler motions, gravitational redshift, and light bending \citep{Fabian:1989ej, Laor:1991nc, Dauser:2010ne}. The radial emissivity profile, often parameterized as a power law  with
$\epsilon(r)\;\propto\;r^{-q}$,
directly traces the corona’s illumination of the disk. General-relativistic ray‑tracing calculations predict a steep inner index ($q\gtrsim3$-5) within $\sim30\,r_g$ {\citep{Wilkins:2012zm, Svoboda:2012cy, Gonzalez:2017gzu}}, which is driven by light‑bending and relativistic focusing, and a transition to a flatter decline --- $q\to3$ {\citep{Zhang:2024ahe}} or even $q\sim2$ for extended corona {\citep{Kinch:2016ipi}} -- at larger radii \citep{Dauser:2013xv, Wilkins:2012zm}. By fitting $q$ via detailed line‑profile modeling, one constrains the corona’s height, radial extent, and geometry (lamp-post vs.\ extended).

Emission from disk radii $r\lesssim6\,r_g$ experiences extreme relativistic effects. Orbital Doppler shifts broaden the line, special-relativistic Doppler boosting enhances the blue wing, and gravitational redshift produces a pronounced red tail. Thus, the precise shape of this profile depends sensitively on the location of the disk’s inner edge, commonly associated with the innermost stable circular orbit (ISCO), as well as on the local orbital velocity field and photon trajectories through curved space-time. Hence, by modeling the iron line profile, one can infer the dimensionless spin parameter, $a_*$ , of the central object under the assumption that the disk truncates at the ISCO \citep{Reynolds:2013rva, Wilkins:2020pgu}. For example, modeling the full transfer function including ray-tracing for photon geodesics in the Kerr metric \citep{Laor:1991nc, Reynolds:2020jwt} yields constraints on disk inclination, $i$, the ionization parameter, $\xi$, and the dimensionless spin parameter, $a_*$ under the assumption that the disk truncates at the ISCO. Spin measurements have been obtained for both AGNs and X‑ray binaries using models such as \texttt{relline} \citep{Dauser:2013xv} and \texttt{kerrconv} \citep{Brenneman:2006hw}. To date, however, most of such analyses have yielded spin measurements for dozens of black hole candidates, under the physical assumptions that the space-time of the central compact object is described by the Kerr metric.

However, the defining feature of a Kerr black hole and its event horizon may not be unique. A growing body of recent work shows that horizonless compact objects can mimic many of the Kerr metric's observational signatures, including quasi-periodic oscillations, accretion dynamics, and gravitational-wave ringdown \citep[e.g.,][]{Cardoso:2019rvt}. Among these exotic alternatives, traversable ``Kerr-like'' wormholes, as an unusual solution to Einstein's equations, occupy a special place. They are smooth, horizonless solutions that join two asymptotically flat regions via a throat, yet reproduce much of the geodesic structure of the Kerr geometry while introducing new phenomenology near the throat region. 
In Kerr-like wormhole models \citep[e.g.,][]{Bueno:2017hyj, Paul:2019trt}, the line element depends not only on the usual mass, $M,$ and spin, $a_*$, but also on additional wormhole parameters (such as throat radius or shape-function coefficients) that govern the location of the photon sphere and the stability of circular orbits. Although wormholes theoretically allow matter and light to pass from one region of space-time to another, not all wormhole solutions are physically traversable. For example, the Schwarzschild wormhole, which is also known as the Einstein-Rosen bridge, is non-traversable; it pinches off too quickly for any signal or object to cross it \citep{Fuller:1962zza}. 
The first physically traversable wormhole geometries were introduced by \cite{Morris:1988cz}, who showed that maintaining an open throat requires exotic matter that violates the standard energy conditions. These ideas were later extended to more speculative scenarios, including closed time-like curves or ``time machines'' \citep{Morris:1988tu}. For a detailed exposition of traversable wormholes and their role in classical and modified gravity, see Visser's comprehensive monograph \citep{Visser:1995cc} and further discussions in \citep{Lemos:2003jb}.

Wormhole geometries generically require exotic stress–energy that violates the standard energy conditions \citep{Morris:1988cz}.  Nonetheless, various proposals aim to circumvent this requirement.  For example, time‑dependent (dynamical) wormhole solutions can momentarily satisfy the energy conditions during certain phases of their evolution \citep{Kar:1995ss}, and wormholes in modified gravity theories ranging from higher‑order curvature corrections to non-minimal scalar‑tensor couplings can support physically reasonable matter without invoking exotic fields \citep{Kanti:2011jz,Harko:2013yb,Mehdizadeh:2015jra,Shaikh:2015oha,Shaikh:2018kfv}.  The issue of wormhole stability remains an active area of research; some backgrounds are stable against linearized metric and field perturbations \citep{Poisson:1995sv,Armendariz-Picon:2002gjc,Bronnikov:2013coa}, while others rapidly collapse or develop singularities \citep{Gonzalez:2008wd,Bronnikov:2012ch,Cuyubamba:2018jdl}.  Despite these open issues, wormholes continue to inspire investigation as compelling testbeds for strong‑gravity phenomena and potential black hole mimickers.

The ability of wormholes to reproduce key black hole signatures was first emphasized by \cite{Damour:2007ap}, who showed that effects such as quasi‑normal mode spectra, accretion behavior, and no‑hair properties can be nearly indistinguishable from those of Kerr black holes. Following the initial Laser Interferometer Gravitational-Wave Observatory (LIGO) detections of gravitational waves \citep{LIGOScientific:2016aoc}, \cite{Cardoso:2016rao} demonstrated that wormholes with thin shells of “phantom” matter at the throat exhibit ringdown signals that are virtually identical to black holes at early times, diverging only in the late‑time tail.  Subsequent work by \cite{Konoplya:2016hmd} found that certain wormhole classes can match or depart from black hole ringdown across all phases. These results underscore the need for precise observational diagnostics, such as detailed iron line spectroscopy, to distinguish true event horizons from horizonless alternatives (see also \citealt{Bambi:2013nla,Nedkova:2013msa,Ohgami:2015nra,Azreg-Ainou:2014dwa,Ohgami:2016iqm,Abdujabbarov:2016efm,Shaikh:2018kfv,Gyulchev:2018fmd,Amir:2018pcu,Jusufi:2017mav,Shaikh:2018oul,Dai:2019mse}).

Beyond the theoretical allure of wormholes, it is crucial to place these models within the broader context of current X-ray astronomy, where multi-messenger constraints are beginning to break long-standing parameter degeneracies. While lamppost geometry is widely used for its simplicity in reflection modeling, its physical validity and that of the underlying space-time is now being rigorously tested by X-ray polarimetry. Pioneering observations by the Imaging X-ray Polarimetry Explorer (IXPE) of stellar-mass black holes like Cygnus X-1 \citep{Krawczynski:2022obw, Steiner:2024aum} and 4U 1630–47 \citep{Ratheesh:2023axn} have provided independent constraints on the coronal extent and disk inclination. These measurements significantly reduce the geometric freedom in spectral fitting, thereby sharpening the utility of reflection features as probes of the metric itself \citep{Mikusincova:2023bon}. Furthermore, the assumption that the accretion disk extends down to the ISCO remains an active topic of debate. The "truncated disk" model suggests that in certain spectral states (such as the hard state of black hole binaries), the thin disk may terminate at a radius that is significantly larger than the ISCO due to evaporation into a hot, advection-dominated flow \citep{Done:2007nc}. Since wormhole parameters primarily manifest their signatures near the throat (and thus near the ISCO), these astrophysical caveats, i.e., polarimetric constraints on the lamppost height and the possibility of disk truncation, are essential considerations for robustly distinguishing between a Kerr black hole and a wormhole mimicker.

In this work, we concentrate on the iron K$\alpha$ line profile produced by a $\delta$-function fluorescence line in the plasma rest frame, which encodes the extreme orbital velocities and light-bending near the compact object. Modifications to the ISCO radius and photon capture cross section in Kerr-like wormholes imprint characteristic shifts in the red wing and asymmetries in the blue horn, which yields potentially observable signatures in high-quality X-ray spectra. 
To model these effects, with the help of a new subroutine to describe the wormhole geodesic structure, we employed two \textsc{XSPEC}\footnote{http://heasarc.gsfc.nasa.gov/docs/xanadu/xspec/index.html} components publicly released by \cite{Mummery:2024znv}: \texttt{SKLINE}, which computes the isolated line profile for a thin, equatorial disk illuminated by a lamppost corona in a Kerr-like wormhole space-time, and \texttt{SKCONV}, a convolution model that applies \texttt{SKLINE} to full rest-frame reflection spectra including lines, edges, and Compton scattering features for a comprehensive relativistic reflection analysis.

Previous work on non‑Kerr iron line modeling has largely explored metrics arising from modified gravity or additional fields. \cite{Zhou:2016koy} modeled the iron line profile in the X-ray reflection spectrum of a thin accretion disk around rotating Ellis wormholes.
\cite{Liu:2018bfx} calculated line profiles in the Janis–Newman–Winicour space-time with a charged, massless scalar field, while \cite{Zhou:2019kwb} examined profiles in modified Kerr metrics permitting $|a_{\bullet}|>1$ and multiple horizons. \cite{Riaz:2022rlx} and \cite{Riaz:2023yde} employed iron line spectral modeling and X-ray observational data to test regular black hole metrics. Moreover, very recently, \cite{Mummery:2024znv} constructed an iron line model for the Kerr naked singularity as an extension of the previous \textsc{XSPEC} model, which only considered the $|a_{\bullet}|< 1$ case.

To search for Kerr‑like wormholes in our Universe, we constructed a custom \textsc{XSPEC} model that couples a fully relativistic accretion‑disk ray‑tracing code to the standard “relxill” reflection framework. For each choice of spin and wormhole parameters, we computed the iron line response of a thin, equatorial disk illuminated by a lamppost corona, and generated a grid of line profiles covering the full parameter space. In addition to probing the possibly existent space-time tunnel, the inclusion of Kerr-like wormholes in fitting procedures can equally well be thought of as stress testing conventional models of accretion. We then embedded these profiles within a multicomponent reflection spectrum to assess how individual iron line shapes contribute to the overall fit. Finally, to assess detectability with current instrumentation, we simulated \textit{NuSTAR}\footnote{https://heasarc.gsfc.nasa.gov/docs/nustar/} observations of representative wormhole and black hole spectra, folding through realistic background and response matrices. By comparing the simulated data to canonical black hole fits, we demonstrated which regions of wormhole parameter space can be constrained or ruled out by present‑day X‑ray missions, and we discuss prospects for next‑generation observatories. 

This paper is structured as follows.
In Sect.~\ref{sec:ironline} we introduce the X-ray line fitting formalism, and describe the fundamental equations. 
In Sect.~\ref{sec:Observed_Line} we analyze the observational appearance of isolated Fe K$\alpha$ line profiles produced in Kerr-like wormhole space-times. Section~\ref{sec:convolved_spectrum} extends this study to a full rest-frame reflection spectrum convolved with the relativistic transfer functions, and  demonstrate how wormhole signatures persist in realistic spectra. 
In Sect.~\ref{sec:simulation_spectrum} we present synthetic \textit{NuSTAR}/FPMA observations generated from these templates and employ the fitting procedures used to compare Kerr and wormhole models. 
Finally, Sect.~\ref{sec:Summary} summarizes our main findings, discusses parameter dependencies and observational caveats, and outlines directions for future work.

\section{Iron line profile modeling}  \label{sec:ironline}

A common feature of the X-ray spectra of astrophysical black hole candidates is the presence of broad and oblique iron lines produced in the inner part of the accretion disk. The specific flux density $\mathcal{F}$ from a target observed by a distant observer can be calculated as follows \citep{Misner:1973prb}:
\bqn
\mathcal{F}(E_\text{\rm obs}) = \int\int I_E (E_{\rm obs})d\Omega,
\eqn
where $E_{\rm obs}$ and $I_E(E_{\rm obs})$ are the observed photo energy and the specific intensity measured at the location of the distant observer, respectively. The $d\Omega$ is the differential element of the solid angle that is subtended on the sky of the observer by a disk area element. Since $I_E/E^3$ is a relativistic
invariant, the specific flow density can be rewritten as\bqn
\mathcal{F}(E_{\rm obs})=\int\int \bm{g}^3 I_E(E_{\rm emit}) d\Omega,
\eqn
where factor $\bm{g}$ is defined as the ratio of the observed energy, $E_{\rm obs}$ , to the emitted energy, $E_{\rm emit}$:
\bqn    
\label{gfactor}
\bm{g}(r)\equiv \frac{E_{\rm obs}}{E_{\rm emit}} = \frac{-(U^\mu p_\mu)_{\rm obs}}{-(U^\nu p_\nu)_{\rm emit}} =\frac{1}{U^t}\Bigg(1+\frac{p_\phi U^\phi}{p_t U^t}\Bigg)^{-1}, 
\eqn
i.e., the combined gravitational and Doppler energy shift (the so-called $\bm{g}$-factor). Here, $p_\phi$ is the angular momentum and $-p_t$ is the emitted photon energy. The azimuthal and time four-velocity
components of the accreting gas are $U^\phi$ and $U^t$  , respectively.
The differential solid angle in which the radiation is emitted for a stationary observer at a large distance $D$ from the source is given by
\bqn
d\Omega = \frac{d\alpha d\beta}{D^2},
\eqn
where $\alpha$ and $\beta$ are photon impact parameters at infinity. 
Therefore, determining the observed specific flux from an astrophysical source breaks down into knowing the specific intensity of the source in its own rest frame and calculating the energy shifts of the photons on their path to the observer.

The rest frame specific intensity can be expressed, in the case where the emergent spectrum is assumed to be only a $\delta$-function, emission line \cite{Dauser:2010ne} as
\bqn
I_E(E_{\rm emit}) = I _{\rm line} \delta(E_{\rm emit}-E_{\rm line})\epsilon(r),
\eqn
where $\epsilon(r)$ is the disk emissivity function. In this case, the observed specific flux is 
\bqn
\mathcal{F}(E_{\rm obs}) = \frac{I _{\rm line}}{D^2}\int\int \bm{g}^3 \epsilon(r) \delta(E_{\rm obs}-gE_{\rm line})d\alpha d\beta.
\eqn

The disk emissivity is expressed as a broken power law in the radius \citep{Dauser:2013xv, Ingram:2019qlb},  
\bqn  
\epsilon(r) \propto  
\begin{cases}  
r^{-q_{\rm in}}\,, & r \leq r_{\rm br}\,, \\  
r^{-q_{\rm out}}\,, & r > r_{\rm br}\,,  
\end{cases}  
\eqn  
where $q_{\rm in}$ and $q_{\rm out}$ are the inner and outer emissivity indices, respectively, and $r_{\rm br}$ is the break radius in our \textsc{XSPEC} thin accretion disk model. 
{While theoretical models and observations often suggest a broken power law profile to account for relativistic focusing (as discussed in Sec. \ref{sec:intro}), for the sake of simplicity and to minimize the number of free parameters in our initial analysis, we adopted a single unbroken power law profile where $q_{\rm in} = q_{\rm out} \equiv q$. In this setup, the disk extends down to the inner radius $r_{\rm in}=\max[r_{\rm th},\,r_{\rm ISCO}]$.} This choice isolates the impact of the wormhole throat and the absence of an event horizon on the iron line profile. Steeper emissivity profiles would only amplify these effects. A more detailed treatment, including double-broken power law emissivities or intra-ISCO emission, lies beyond the scope of this work and is addressed in our conclusions. 

\begin{figure}
    \centering
    \includegraphics[width=0.9\linewidth]{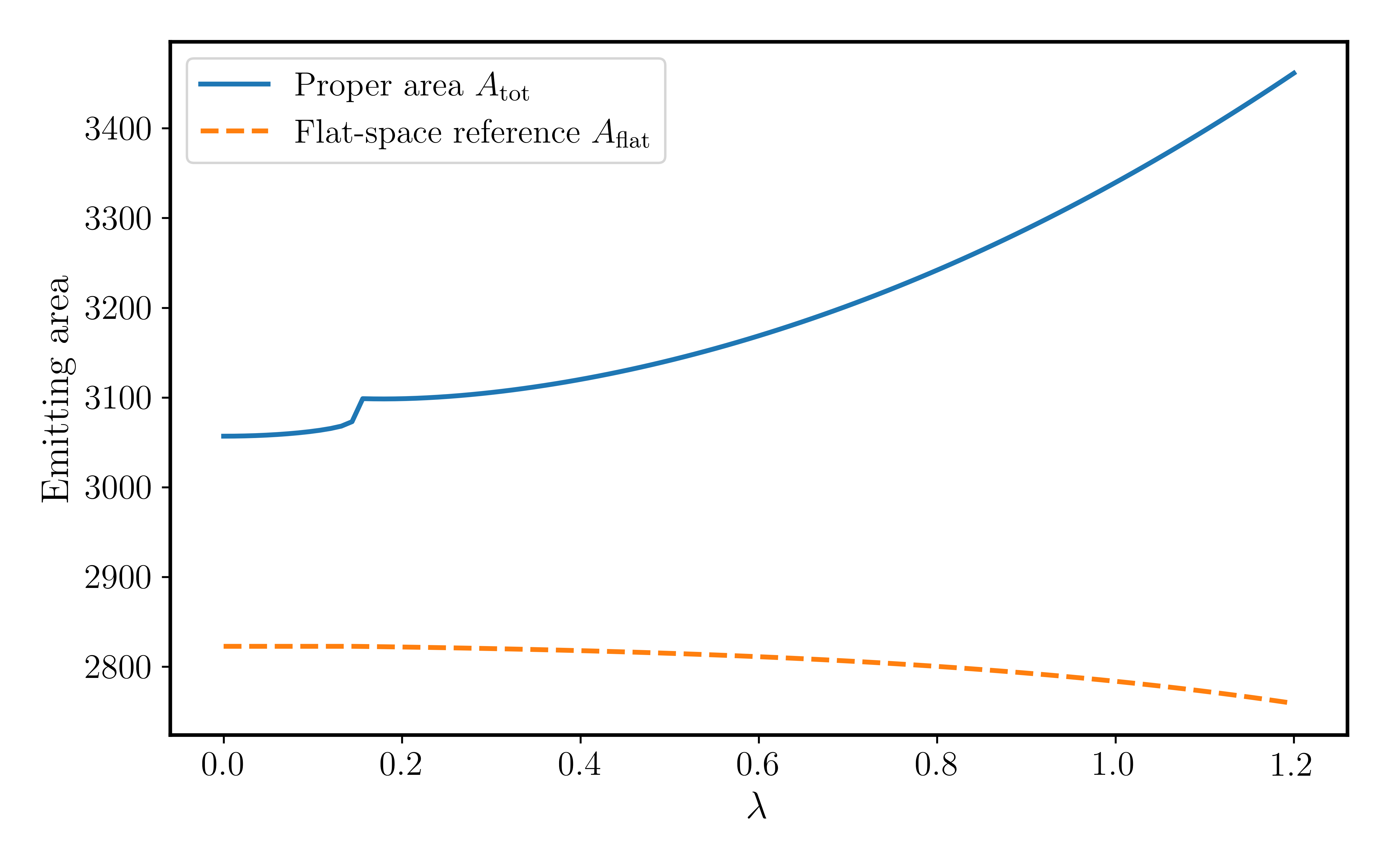}
    \caption{{Total proper emitting area of the equatorial thin disk between $r_{\rm in}=\max[r_{\rm th},\,r_{\rm ISCO}]$ and $r_{\rm out}=30M$ as a function of the wormhole deformation parameter $\lambda$ (spin $a_\ast=0.998$). The solid curve shows the relativistic proper area computed from the disk surface metric, while the dashed curve shows the corresponding Euclidean reference area $\pi(r_{\rm out}^2-r_{\rm in}^2)$ using the same $r_{\rm in}$.}}
    \label{fig:disk_area}
\end{figure}

To quantify how the metric coefficient $g_{rr}$ modifies the effective emitting area of the disk, we computed the proper area of the equatorial thin-disk surface. On the spatial slice $t=\mathrm{const}$ and in the equatorial plane $\theta=\pi/2$, the induced two-metric on the disk is $h_{ab}=\mathrm{diag}(g_{rr},\,g_{\phi\phi})$ (in Boyer--Lindquist--like coordinates where $g_{r\phi}=0$), yielding the proper area element
\bqn
dA=\sqrt{\det(h)}\,dr\,d\phi=\sqrt{g_{rr}g_{\phi\phi}}\,dr\,d\phi.
\eqn
Therefore, the proper area of a narrow annulus is $dA_{\rm disk}=2\pi\sqrt{g_{rr}g_{\phi\phi}}\,dr$, and the cumulative emitting area from the disk inner edge to radius $r$ is
\bqn
A=\int_{r_{\rm in}}^{r}2\pi\sqrt{g_{rr}g_{\phi\phi}}\,dr.
\eqn
For the comparison, we fixed $r_{\rm out}=30M$ and evaluated $A_{\rm tot}(\lambda)=A(<r_{\rm out})$ for $\lambda\in[0,1.2]$, adopting the disk inner edge $r_{\rm in}(\lambda)=\max[r_{\rm th}(\lambda),\,r_{\rm ISCO}(a_\ast)]$ to enforce that the emitting region starts outside both the throat and the ISCO.

{Figure~\ref{fig:disk_area} shows that the proper emitting area increases systematically with $\lambda$ for a fixed $r_{\rm out}$, demonstrating that the wormhole deformation alters the mapping between the coordinate radius and the physical surface element through $\sqrt{g_{rr}}$ (and the associated $g_{\phi\phi}$ contribution). For reference, we also ploted the Euclidean disk area $A_{\rm flat}=\pi(r_{\rm out}^2-r_{\rm in}^2)$ computed using the same $r_{\rm in}(\lambda)$. The separation between $A_{\rm tot}$ and $A_{\rm flat}$ highlights the genuine relativistic correction to the emitting area driven by the curved geometry. A distinct change in the trend occurs when $r_{\rm th}(\lambda)$ surpasses $r_{\rm ISCO}$, at which point the inner edge becomes throat-limited rather than ISCO-limited, producing a {small discontinuous} feature associated with the switch in $r_{\rm in}(\lambda)$. Overall, this diagnostic explicitly visualizes the influence of $g_{rr}$ on the cumulative emitting area at each radius.}

\begin{figure}
    \centering
    \includegraphics[width=0.85\linewidth]{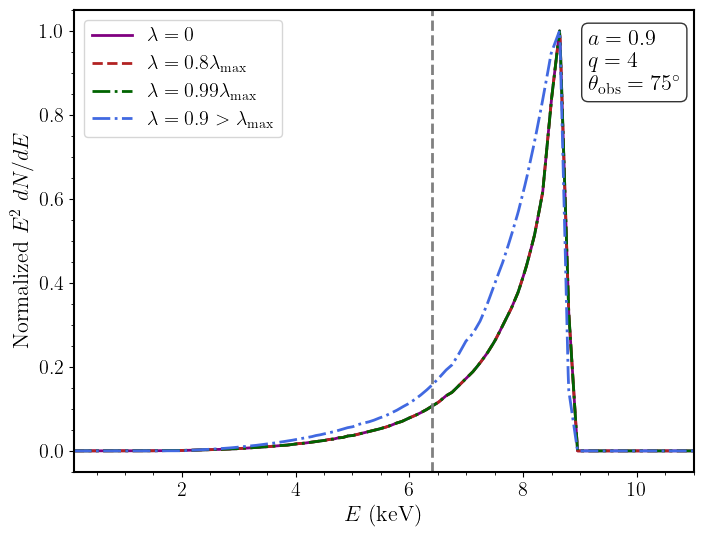}
   \caption{Synthetic Fe K$\alpha$ line profiles for the Kerr black hole ($\lambda = 0$, solid purple line) and Kerr-like wormholes with various deformation parameters, $\lambda$ , at spin $a = 0.9$. The dashed red and dash-dotted green  lines represent wormholes with $\lambda = 0.8\lambda_{\rm max}$ and $\lambda = 0.99\lambda_{\rm max}$, respectively, demonstrating a near-perfect mimicking effect when the throat, $r_{\rm th}$ , is within the ISCO. In contrast, the dash-dotted blue line shows a significant spectral deviation when the wormhole throat extends beyond the ISCO ($\lambda = 0.9 > \lambda_{\rm max}$). All curves are normalized and plotted as $E^2\,\mathrm{d}N/\mathrm{d}E$. The model parameters are observer inclination angle $\theta_{\rm obs}=75^\circ$, emissivity index $q=4$, rest-frame line energy $E_{\rm line}=6.4\ \mathrm{keV}$ (vertical dashed line), outer disk radius $r_{\rm out}=30\,r_g$, and disk surface density $\mathrm{Dens}=16$.}
    \label{fig:mimicker}
\end{figure}

\begin{figure}
    \centering
    \includegraphics[width=0.85\linewidth]{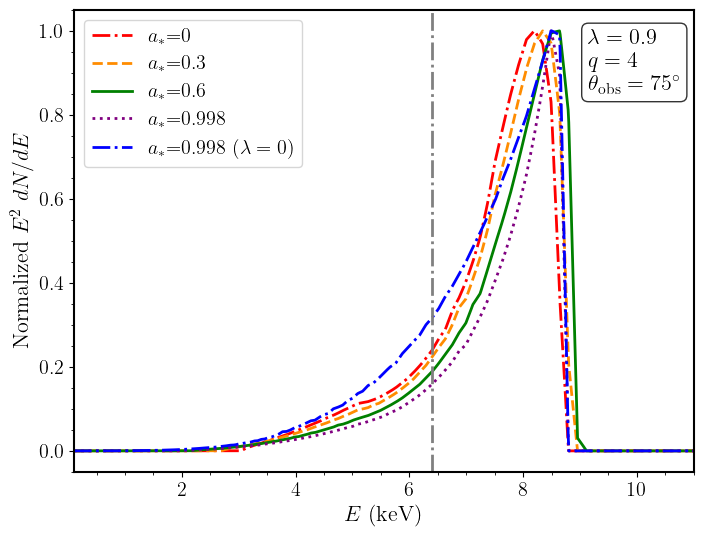}
    \caption{Synthetic Fe K$\alpha$ line profiles from a geometrically thin disk in a Kerr black hole (dashed blue lines) and Kerr-like wormhole with the deformation parameter $\lambda=0.9$ (solid lines). Different colors represent different spins, $a_*=0.3$ (red), $0.6$ (green), and $0.998$ (purple).
    All curves are normalized and plotted as $E^2\,\mathrm{d}N/\mathrm{d}E$. Other model parameters are the same as in Fig. \ref{fig:mimicker}. 
}
    \label{KerrWH_Feline1}
\end{figure}

\begin{figure}
    \centering
    \includegraphics[width=0.85\linewidth]{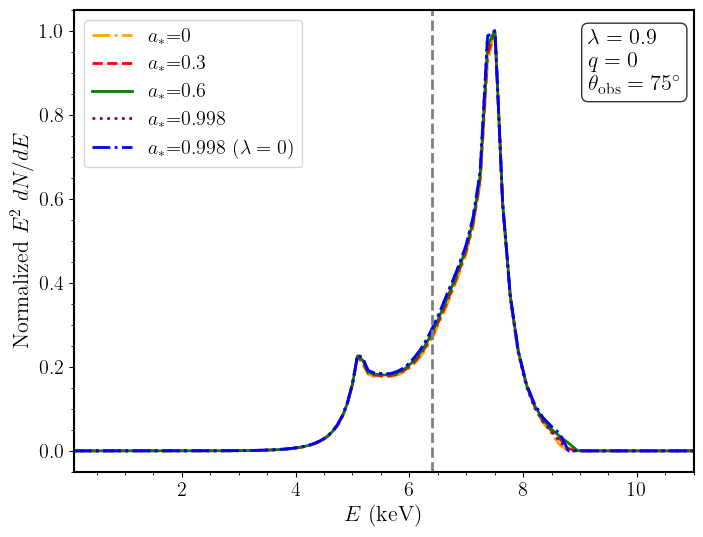}
    \caption{Synthetic Fe K$\alpha$ line profiles from a
    geometrically thin disk with $q=0$ (i.e., constant emissivity per unit area) in a Kerr-like wormhole with the deformation parameter $\lambda=0.9$. The different colors represent different spins, $a_*=0$ (dash-dotted~orange), $0.3$ (dashed red), $0.6$ (solid green), and $0.998$ (dotted purple). The dash-dotted blue line represents the case of a Kerr black hole ($\lambda=0$) with $a_*=0.998$. Other model parameters are the same as those in Fig.~\ref{KerrWH_Feline1}. 
}
    \label{KerrWH_Feline2}
\end{figure}

\begin{figure}
    \centering
    \includegraphics[width=0.85\linewidth]{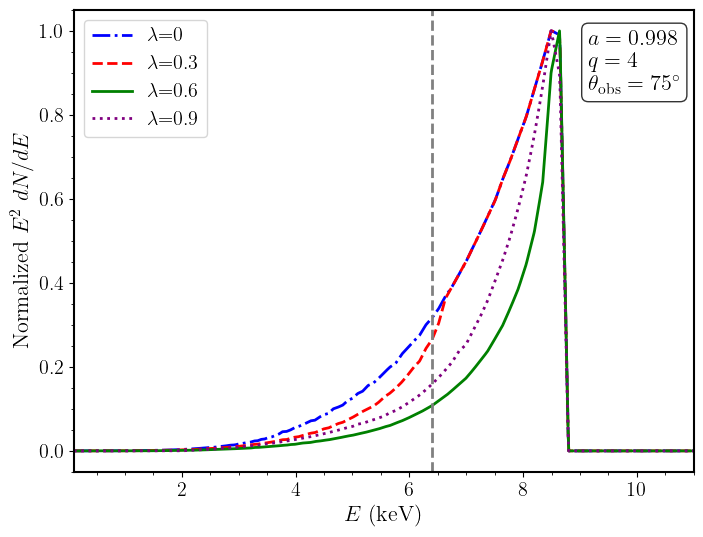}
    \caption{Same as Fig.~\ref{KerrWH_Feline1} but using different deformation parameters: $\lambda=0$ (dash-dotted~blue), $0.3$ {dashed red}, $0.6$ (solid~green), and $0.9$ (dotted purple) with a fixed spin $a_*=0.998$.
}
    \label{KerrWH_Feline3}
\end{figure}

\begin{figure}
    \centering
    \includegraphics[width=0.85\linewidth]{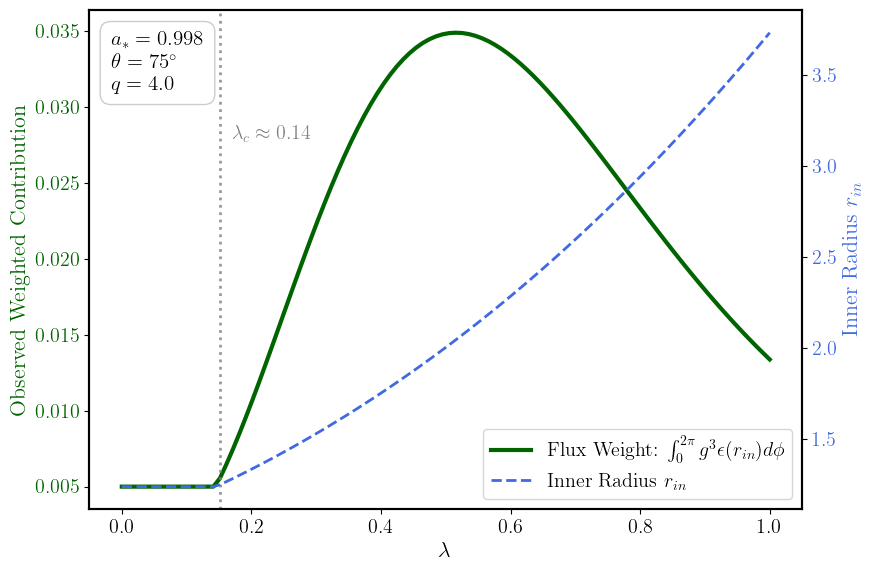}
\caption{{Dependence of the inner-radius emission contribution on the wormhole parameter $\lambda$. The solid green line represents the observed weighted flux $\int_{0}^{2\pi} \bm{g}^3 \epsilon(r_{in}) d\phi$, while the dashed blue line indicates the evolution of the inner disk radius, $r_{in}$. The vertical dotted line marks the critical value $\lambda_c \approx 0.14$, where the wormhole throat exceeds the ISCO radius. For $\lambda > \lambda_c$, the observed intensity exhibits a non-monotonic behavior due to the interplay between relativistic Doppler boosting and the radial dilution of the emissivity, $\epsilon(r)$. The model parameters are $a_* = 0.998$, $\theta = 75^\circ$, and $q = 4.0$.}}    \label{fig:gfactor}
\end{figure}

\begin{figure}
    \centering
    \includegraphics[width=0.85\linewidth]{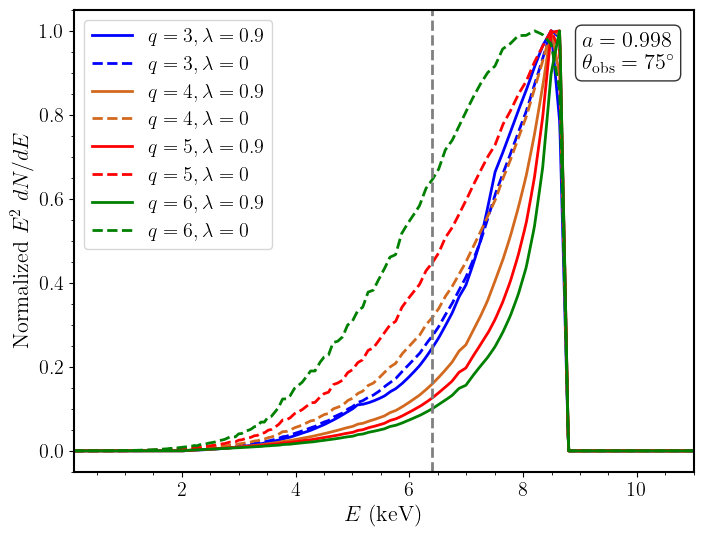}
    \caption{Same as Fig.~\ref{KerrWH_Feline1} but using different emissivity parameters: $q=3$ (blue), $4$ (orange), $5$ (red), and $6$ (greed). The solid lines show a Kerr-like wormhole case ($\lambda=0.9$), and the dashed lines show a Kerr black hole case ($\lambda=0$) with a fixed spin $a_*=0.998$.
    }
    \label{KerrWH_Feline4}
\end{figure}

\begin{figure}
    \centering
    \includegraphics[width=0.85\linewidth]{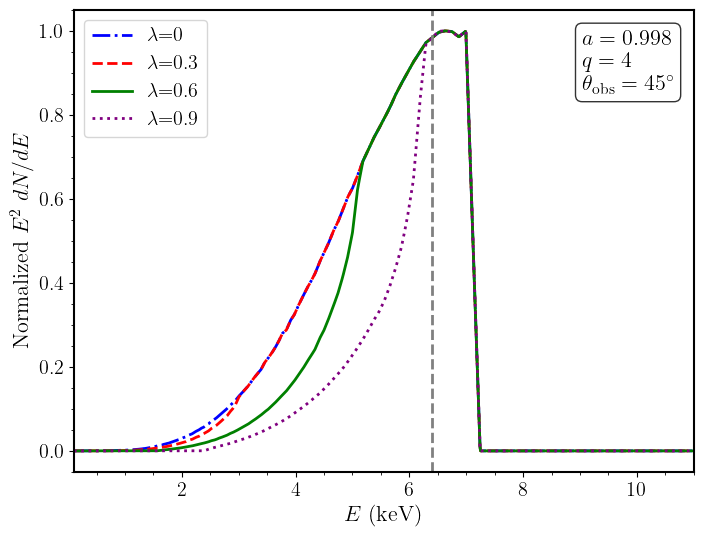}
    \caption{Same as Fig.~\ref{KerrWH_Feline3} but using a different inclination angle, $\theta_{\rm obs}=45^\circ$.  
    }

    \label{KerrWH_Feline5}
\end{figure}

\begin{figure}
    \centering
    \includegraphics[width=0.85\linewidth]{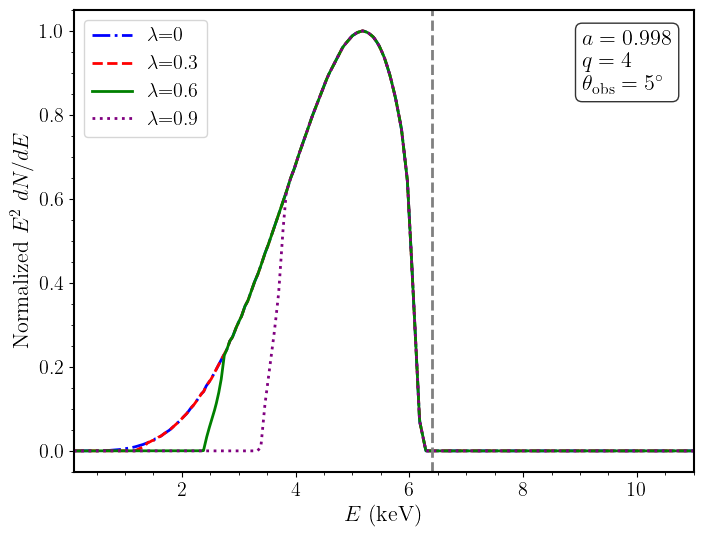}
    \caption{Same as Fig.~\ref{KerrWH_Feline3} but using a different inclination angle, $\theta_{\rm obs}=5^\circ$.
    }

    \label{KerrWH_Feline6}
\end{figure}

{To construct the observed profiles, we placed a Cartesian grid $(\alpha, \beta)$ on the observer's image plane and integrate the null geodesics backward to the accretion disk. As described in Sec.~\ref{subsec:nullgeodesics}, our implementation incorporates the non-Kerr radial geometry by applying a radial integral correction to the Weierstrass elliptic integrals and redefining the horizon-dependent poles to match the wormhole throat radii $r_\pm$. Specially, within the \texttt{INTRPART} and \texttt{YNOGK} modules, we applied a radial integral correction to the Weierstrass elliptic integrals. These customizations ensure that the calculation of the Boyer-Lindquist coordinates $(r, \mu, \phi, t)$ and the affine parameter $\sigma$ accurately account for the coordinate time-delay and the radial metric deformation introduced by the wormhole parameter $\lambda$. 
Substituting the resulting interception radii and redshift factors into the surface integral yields the observed line profiles shown in the following sections.}

The observed isotropic luminosity at energy $E_{\rm obs}$ is obtained by numerically evaluating the surface integral as 
\bqn  
L_{E} = 4\pi \iint_{S} \bm{g}^3 f_{\rm col}^{4} B_{\rm E}\!\left(E_{\rm obs}/\bm{g},\, f_{\rm col}T(r_f)\right) \mathrm{d}\alpha\,\mathrm{d}\beta,  
\eqn  
where $f_{\rm col}$ is the color temperature correction, $T(r_f)$ is the local disk temperature at the photon interception radius $r_f$, $B_{\rm E}$ is the blackbody spectrum, and the energy shift, $\bm{g}$, is determined by the disk fluid's 4-velocity and the photon's initial conditions as defined in Eq. \eqref{gfactor}.
To construct the image, we placed a Cartesian grid $(\alpha,\beta)$ on a plane perpendicular to the line of sight at inclination $i$ and distance $D$. Each point maps to spherical coordinates $(r_{i},\theta_{i},\phi_{i})$ via  
\begin{align}  
r_{i} &= \sqrt{D^{2} + \alpha^{2} + \beta^{2}}, \\  
\cos\theta_{i} &= \frac{D\cos i + \beta\sin i}{r_{i}}, \\  
\tan\phi_{i} &= \frac{\alpha}{D\sin i - \beta\cos i}.  
\end{align}  
Imposing orthogonality between the photon momentum and the image plane fixes the initial 4-velocity components $(u_{i}^{r},u_{i}^{\theta},u_{i}^{\phi})$ (see \citealt{Yang:2013yoa,Dexter:2009fg}). We then integrated the Kerr-like wormhole null geodesics backward from each $(\alpha,\beta)$ to the disk plane using \texttt{YNOGK}, recording the interception radius $r_f$ and $p_{\phi}/p_{t}$. Substituting these into the surface integral yields the observed line profile.

\section{{Synthetic Fe line profiles}}  
\label{sec:Observed_Line}

In this section we present ray-traced calculations of relativistically broadened iron line profiles in Kerr-like wormhole space-times, {computed from geometrically thin, optically thick accretion disks extending from an inner radius $r_{\rm in} = \max(r_{\rm th}, r_{\rm ISCO})$ to $r_{\rm out} = 30r_g$, with rest‑frame line energy $E_{\rm line}=6.4,\mathrm{keV}$ and observer inclination angles from $5^\circ$ to $75^\circ$. Emission is normalized as $E^2\mathrm{d}N/\mathrm{d}E$ and follows a power law emissivity $\epsilon(r)\propto r^{-q}$.} Our analysis focuses on distinguishing features between Kerr-like wormhole and Kerr black hole spectra, which may serve as observational signatures of exotic compact objects.

The structure of the image plane in a Kerr-like wormhole space-time differs slightly from that of a black hole, leading to pronounced differences in the resulting iron line profiles across most of the parameter space. 
However, such differences significantly diminish in specific regimes. As illustrated in Fig.~\ref{fig:mimicker}, when the deformation parameter $\lambda$ is below the critical threshold $\lambda_{\rm max}$, the resulting iron line profiles (dashed red and dash-dotted green lines) become nearly indistinguishable from the Kerr black hole case. Since the wormhole throat $r_{\rm th}$ is located inside the ISCO radius, the ISCO acts as a shield, ensuring that the observed flux is dominated by the accretion disk in a region where the metric closely resembles that of a Kerr black hole. Significant spectral deviations only emerge when $\lambda > \lambda_{\rm max}$ (dash-dotted blue line), in which case the throat extends beyond the ISCO and modifies the emitting region of the disk.

Beyond this mimicker regime, the distinct geometric features of the wormhole become more apparent. Figure~\ref{KerrWH_Feline1} presents the computed line profiles for a steep emissivity gradient ($q=4$), where the radiation is predominantly concentrated in the innermost region of the accretion disk. In this regime, the spectral shape is sensitive to the geometric properties of the central compact object. For a spin parameter $a = 0.998$, the Kerr black hole case ($\lambda=0$) exhibits a more pronounced redshifted tail than the Kerr-like wormhole case ($\lambda=0.9$, dotted purple line), the latter's specific parameters configuration are summarized in Table~\ref{tab:kwline}, indicating that at the same near-extremal spin, the gravitational redshift near the black hole contributes more strongly to the low-energy flux. In addition, the blueshifted peak in the wormhole model is shifted slightly toward lower energies compared to the black hole model. As the spin increases from $a=0.3$ to $a=0.998$, the red wing in both models extends significantly toward lower energies, reflecting that the inner disk edge moves closer to the central object with increasing spin and thus probes a deeper gravitational potential well.

{However, when the emissivity index is reduced to $q=0$ (as shown in Fig.~\ref{KerrWH_Feline2}), the emissivity becomes uniform across the disk surface, and the contribution from the outer disk substantially dilutes the relativistic effects produced in the strong-gravity region. In this case, the spectrum transitions from a single-peaked, skewed profile to the characteristic double-peaked shape dominated primarily by Doppler boosting. Most importantly, under $q=0$, the spectral differences among different spin parameters $a_*$ nearly vanish, and the Kerr-like wormhole and Kerr black hole profiles become difficult to distinguish. This demonstrates that the discriminating power of the iron line profile depends on the disk's physical properties: only when most of the emission is released from the region close to the central object can X-ray spectral observations effectively identify Kerr-like wormholes and differentiate them from conventional black holes.}

{The evolution of the iron line profiles with respect to the deformation parameter $\lambda$ reveals a complex, non-monotonic relationship that challenges simple geometric expectations. As shown in Fig.~\ref{KerrWH_Feline3}, for a fixed spin of $a_* = 0.998$, the Kerr black hole ($\lambda=0$, dash-dotted blue line) exhibits the most extended red wing, reaching down to approximately 2 keV. As the space-time deviates from the Kerr metric, the red wing progressively narrows; however, the profile for $\lambda = 0.6$ (solid green line) emerges as the narrowest among the cases studied, appearing even more truncated than the extreme wormhole case of $\lambda = 0.9$ (dotted purple line). This suggests that the observational distinction between these objects does not follow a linear trend with the deformation parameter.}

{This peculiar spectral behavior is physically rooted in the interplay between the disk’s inner truncation and the observed flux weighting. According to the "dependence of inner-radius emission" analysis, the inner radius $r_{\rm in}$ (dashed blue line) remains stable for small deformations ($0 \le \lambda < 0.14$) but increases sharply thereafter as the wormhole throat geometry begins to dominate. When $\lambda$ reaches 0.6, the disk is truncated at a significantly larger radius ($r_{\rm in} \approx 2.5 M$), effectively removing the highly redshifted emission from the deepest regions of the potential. While a further increase to $\lambda = 0.9$ pushes the disk even farther out ($r_{\rm in} \approx 3.32 r_g$), the spectral narrowing is offset by a sharp decline in the Observed Weighted Contribution (solid~green curve), which modulates the amplitude and slope of the resulting red wing.
Consequently, the minimum width observed at $\lambda = 0.6$ represents a critical point where the suppression of gravitational redshift due to the retreating inner edge reaches its maximal observational impact relative to the normalized peak. For $\lambda > 0.6$, although the physical redshift at $r_{\rm in}$ continues to decrease, the rapid drop in the flux weight at the inner rim changes the profile's shape, causing the $\lambda = 0.9$ line to appear slightly flatter or less sharply truncated in its decay. This result highlights that Fe K$\alpha$ spectroscopy is sensitive to a specific window of deformation, where the relativistic effects of the wormhole throat are most distinctively imprinted on the observed X-ray spectrum.}

{The concentration of disk emission, characterized by the emissivity index $q$, plays a decisive role in the observational distinguishability of exotic compact objects. As shown in Fig.~\ref{KerrWH_Feline4}, increasing $q$ from 3 to 6 significantly broadens the iron line profiles for both the Kerr black hole and the Kerr-like wormhole. This broadening occurs because a steeper emissivity profile concentrates radiation in the innermost regions of the disk, where relativistic effects, specifically gravitational redshift and Doppler boosting, are most extreme. Notably, the spectral discrepancy between the black hole ($\lambda=0$) and the wormhole ($\lambda=0.9$) becomes more pronounced at higher $q$. For $q=6$, the Kerr black hole exhibits a vastly more extended red wing compared to the wormhole, whose profile remains truncated due to the retreating inner disk edge. This suggests that the distinct space-time geometry of a wormhole is only clearly imprinted on the spectrum when the accretion physics concentrates emission near the throat.
}

{The influence of the observer's inclination angle $\theta_{\rm obs}$ further modifies these spectral signatures, as demonstrated in Figs.~\ref{KerrWH_Feline5} and \ref{KerrWH_Feline6}. At an intermediate inclination of $\theta_{\rm obs} = 45^\circ$, the non-monotonic evolution of the line width with respect to $\lambda$ remains a robust feature. The $\lambda=0.6$ profile (solid green line) continues to appear as the narrowest case, confirming that the interplay between the shifting ISCO and flux weighting is independent of extreme viewing angles. However, as the inclination decreases, the Doppler-induced broadening diminishes, leading to a narrowing of all profiles and a reduction in the blue-shifted peak's intensity compared to high-inclination observations.
}

{In the near face-on limit ($\theta_{\rm obs} = 5^\circ$), the iron line profiles undergo a significant transformation, as shown in Fig.~\ref{KerrWH_Feline6}. The rotational velocity of the disk is nearly perpendicular to the line of sight, effectively eliminating Doppler boosting and leaving the spectrum to be dominated by gravitational redshift and the transverse Doppler effect. Under these conditions, a strong observational degeneracy emerges between the Kerr black hole and wormholes with low deformation parameters ($\lambda = 0.3$), as their profiles almost perfectly overlap. Distinguishable features only reappear for high deformation values ($\lambda \ge 0.6$), where the profiles shift toward higher energies due to the outward migration of the disk's inner boundary. These results collectively indicate that the most favorable conditions for identifying Kerr-like wormholes via X-ray spectroscopy are high-inclination systems with disk emission concentrated near the central object.
}

{The simulated Fe K$\alpha$ line profiles demonstrate that while Kerr-like wormholes are {potential} mimickers of Kerr black holes, they exhibit distinguishable spectral signatures across a wide range of the investigated parameter space. Our calculations reveal that wormhole spectra are generally characterized by narrower line widths, typically reduced by $\sim 20-40\%$ compared to Kerr results, and a significant truncation of the red wing due to the geometric properties of the wormhole throat. While these exotic signatures are most pronounced in systems with high inclination ($\theta_{\rm obs} = 75^\circ$) and steep emissivity ($q \ge 4$), they remain identifiable at moderate inclinations ($\theta_{\rm obs} = 45^\circ$). However, our analysis also uncovers specific regimes of observational degeneracy, such as in cases of flat emissivity ($q=0$) or near face-on viewing angles ($\theta_{\rm obs} = 5^\circ$), where the distinctive space-time features are largely suppressed. Despite these degeneracies, the unique combination of a narrower profile with a weakened red wing at high spin remains a robust indicator of wormhole geometry. Such features are within the sensitivity limits of current missions like \textit{XRISM} and could be definitively resolved by next-generation observatories such as \textit{Athena} or \textit{eXTP}. These findings suggest that high-resolution X-ray spectroscopy provides a viable "smoking gun" signature to discriminate between standard black holes and exotic compact objects.}

\section{Reflected X-ray spectrum}
\label{sec:convolved_spectrum}

Modern reflection spectroscopy constructs the observable X-ray spectrum of an accreting compact object by convolving physically computed rest-frame reflection spectra with a relativistic transfer function that encodes Doppler shifts, gravitational redshifts, and light-bending \citep{Reynolds:2020jwt}. In practice, this approach separates the problem into two pieces: (i) a detailed microphysical rest-frame model of the disk surface (for example, the \texttt{xillver} family), which determines the intrinsic line shapes and the reflected continuum as a function of local ionization, density, and composition; and (ii) a relativistic blurring kernel computed from the space–time of central object and the disk emissivity, which maps local emission at each disk radius and azimuth to what a distant observer sees. The observable profile therefore depends on a set of parameters: the observer inclination $\theta_{\rm obs}$, the inner and outer disk radii $r_{\rm in}$, $r_{\rm out}$ (or the effective ISCO), the radial emissivity (or coronal source geometry, e.g., height $b_X$ in units of $r_g$), the ionization parameter $\xi$, elemental abundances (commonly expressed as iron abundance $A_{\rm Fe}$), the irradiating continuum shape (photon index $\Gamma$), the disk density, and the space-time parameters (spin $a$ or any deformation parameters that distinguish a Kerr-like wormhole from a Kerr black hole). Note that the absolute black hole mass cancels out when distances are expressed in gravitational radii $r_g$, thus the spectral line shapes are insensitive to mass scaling. In the sections that follow, we applied this pipeline by taking \texttt{xillver} rest-frame outputs and convolving them with relativistic transfer functions computed for our Kerr-like wormhole metric, producing the predicted, observer-frame iron K$\alpha$ line profiles and full reflection spectra to compare directly with the standard Kerr black hole case.

We began with a generic rest-frame spectrum $I_E(E)$ emitted at each disk location $(r,\phi)$, which was assumed to be separable into a location-independent shape and a location-dependent emissivity $\epsilon(r)$. Under this factorization, the rest-frame emission could be expressed as:
\bqn
\bm{I_E}(E,r) = \epsilon(r)\int_0^\infty I_E(E')\,\delta(E-E')\,\mathrm{d}E',
\eqn
and the observed flux becomes
\bqn
F_E \propto \iint \bm{g}^3\,\epsilon(r)\int_0^\infty I_E(E')\,\delta\bigl(E_{\rm obs}/\bm{g} - E'\bigr)\,\mathrm{d}E'\,\mathrm{d}\Omega.
\eqn
Reordering integrations yields a convolution form
\bqn
F_E \propto \int_0^\infty I_E(E')\left[\iint \bm{g}^3\,\epsilon(r)\,\delta\bigl(E_{\rm obs}/\bm{g} - E'\bigr)\,\mathrm{d}\Omega\right]\mathrm{d}E',
\eqn
where the bracketed term is the relativistic transfer function (or Green's function) mapping rest-frame energy $E'$ to observed energy $E_{\rm obs}$ (see \citealt{Ingram:2019qlb}).

To generate the realistic reflection spectra presented in Figs.~\ref{Xspectra_a0998} and \ref{Xspectra_lambda09}, we performed a convolution of the general relativistic effects with the rest-frame reflection model using \textsc{XSPEC}. We employed $\chi^2$ statistics with Levenberg–Marquardt minimization, adopting \texttt{angr} abundances and \texttt{vern} cross sections. {The cosmology was fixed to $H_0=70~\mathrm{km\,s^{-1}\,Mpc^{-1}}$, $\Omega_M=0.27$, and $\Omega_\Lambda=0.73$.} The energy grid spans 0.05–100~keV with 12,000 logarithmically spaced channels, providing a high resolution of $\Delta E/E\sim6\times10^{-4}$. 
The total observed spectrum is constructed using the model
\bqn
   \texttt{kwconv*atable\{xillverCp\_v3.6.fits\}},\nb
\eqn
with {\tt kwconv} parameters set to spin $a_*=0.998$, inclination angle $\theta_{\rm obs}=75^\circ$, inner and outer emissivity indices $q_{\rm in}=q_{\rm out}=4$, break radius $r_{\rm br}=15~r_g$, inner radius $r_{\rm in}=1.237,r_g$, outer radius $r_{\rm out}=30\,r_g$, and deformation $\lambda=0, 0.9$.  The {\tt xillver\_Cp} table model uses photon index of the illuminating power law spectrum $\Gamma=2.0$, iron abundance $A_{\rm Fe}=1.0$ with respect to the solar value, ionization $\log\xi=3.1$, high-energy cutoff $E_{\rm cut}=300$~keV, electron density $n_e=10^{16}~\mathrm{cm^{-3}}$, inclination angle $75^\circ$, and normalization $2.6\times10^{-5}$. These are listed in the Table~\ref{tab:xspec-params}.

{Figure~\ref{Xspectra_a0998} compares the convolved spectra for a maximally rotating source ($a_*=0.998$) in two distinct space-times: a Kerr black hole ($\lambda=0$, solid green line) and a Kerr-like wormhole ($\lambda=0.9$, dashed red line). The dotted blue line represents the rest-frame \texttt{xillver} spectrum before relativistic blurring. The comparison highlights that the wormhole geometry produces a less severe "smearing" of the spectral features compared to the black hole. Specifically, the Fe K$\alpha$ emission line in the wormhole spectrum retains a higher peak intensity and a sharper profile, whereas the black hole spectrum exhibits a more extended red wing and a flatter peak. This confirms our previous monochromatic line analysis that the truncated red wing inherent to the wormhole metric manifests in the full reflection spectrum as a suppression of the extreme gravitational broadening typically associated with the ISCO of a Kerr black hole. Figure~\ref{Xspectra_lambda09} presents the convolved reflection spectra for a Kerr-like wormhole with a fixed deformation parameter $\lambda=0.9$, comparing a non-spinning case ($a_*=0$, solid green line) against a high-spin case ($a_*=0.9$, dashed red line). In striking contrast to the black hole comparison in Fig.~\ref{Xspectra_a0998}, the two profiles in Fig.~\ref{Xspectra_lambda09} exhibit a high degree of similarity, appearing nearly indistinguishable in the logarithmic plot. This phenomenon reveals a significant observational degeneracy between spin states in highly deformed wormhole space-times. Physically, this convergence is driven by the geometric truncation of the accretion disk; as demonstrated in the inner-radius analysis, a large $\lambda$ value pushes the disk's inner edge $r_{\rm in}$ outward to approximately $3.32 r_g$, effectively isolating the emitting matter from the deepest regions of the gravitational potential where spin-dependent relativistic effects are most prominent. Furthermore, the observed weighted contribution from the inner disk is substantially diluted at $\lambda=0.9$, meaning the spectral signatures typically used to diagnose spin, such as extreme Doppler broadening and gravitational smearing, are heavily suppressed by the background geometry. Consequently, the reflection features, including the Fe K$\alpha$ complex and the Compton hump, become dominated by the more stable physical conditions of the outer disk regions. These findings suggest that for exotic compact objects with significant deformation, the space-time geometry $\lambda$ can effectively mask the spin $a_*$, posing a challenge for parameter estimation using standard Kerr-based reflection models.

These examples underscore that, even without modeling radial variations in ionization or density, the convolution of standard reflection spectra with relativistic transfer functions is sufficient to discriminate between a Kerr black hole and a Kerr-like wormhole. Incorporating more detailed rest-frame physics (e.g., radius-dependent ionization) will further enhance diagnostic possibilities and is left for future work.

\begin{figure}
    \centering
    \includegraphics[width=1.0\linewidth]{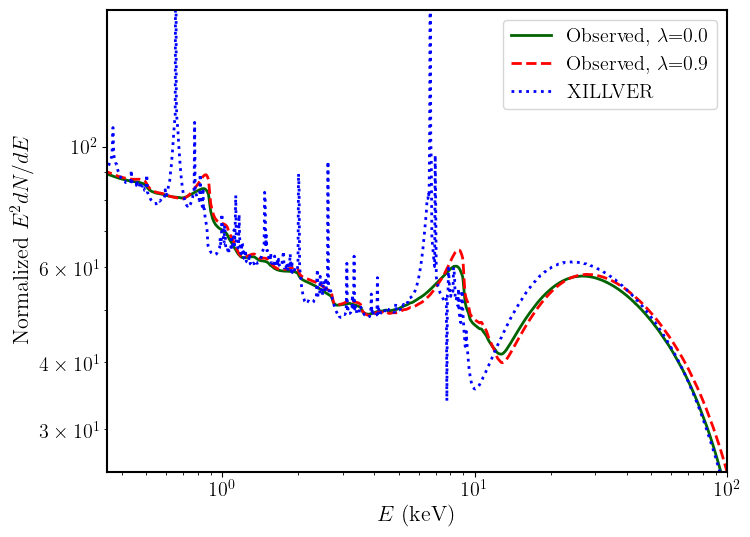}
    \caption{Convolved X‑ray reflection spectrum from a Kerr‑like wormhole with the same spin value $a_*=0.998$ for a Kerr black hole ($\lambda=0.0$, solid green line) and a Kerr-like wormhole ($\lambda=0.9$, dashed red line). The underlying rest‑frame spectrum (dotted blue) was generated with the \texttt{xillver} model (v3.6).
    Detailed parameters are listed in Table~\ref{tab:xspec-params}. 
}
    \label{Xspectra_a0998}
\end{figure}

\section{Simulated \textit{NuSTAR} spectra}
\label{sec:simulation_spectrum}

\begin{figure}
    \centering
    \includegraphics[width=1.0\linewidth]{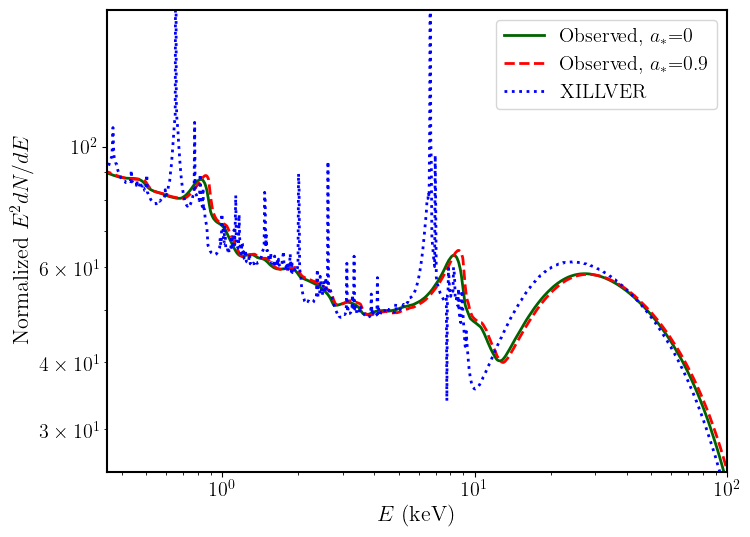}
    \caption{Same as Fig.~\ref{Xspectra_a0998} but showing a Kerr-like wormhole with $\lambda=0.9$ for spin parameters $a_*=0.0$ (solid green line) and $a_*=0.9$ (dashed red line).
    }
    \label{Xspectra_lambda09}
\end{figure}

In this section we describe two sets of \textit{NuSTAR} simulations designed to validate our spectral–fitting procedure on a bright X‑ray binary and explore the feasibility of distinguishing Kerr–like wormholes from Kerr black holes using iron‐line diagnostics.

\subsection{Synthetic bright X‑ray binary spectra}
\label{ssec:binary}

We simulated a 50~ks \textit{NuSTAR} observation of a bright X‑ray binary by adopting the model parameters listed in Table~\ref{tab:xspec-params} within the \textsc{XSPEC} framework. The continuum normalization is adjusted to yield an intrinsic 4–10~keV flux of $5\times10^{-9}\,\mathrm{erg\,cm^{-2}\,s^{-1}}$ with a photon index $\Gamma\simeq2$, matching typical values measured for the bright hard state of GX\,339–4 \citep{Chand:2024isa}. We produce a single FPMA exposure using \textsc{XSPEC}’s \texttt{fakeit} command, including the standard background spectrum but neglecting detector dead‐time; omitting FPMB is formally equivalent to merging both modules with an overall dead‐time correction of 0.5. This configuration replicates a high‐quality observation commonly employed in spectral analyses \citep[e.g.,][]{Parker:2015fja,Tomsick:2018ywi,Xu:2017yrm}.

We loaded the \textit{NuSTAR} RMF and ARF files, which map detector channels to photon energies and encode the effective area, respectively. The background spectrum (\texttt{bgd\_60arcsec.pha}) reproduces the telescope’s intrinsic detector and cosmic background under identical extraction conditions. Both source and background spectra are assigned an exposure time of 50~ks, and the correction normalization is set to unity for nominal calibration.
Prior to invoking \texttt{fakeit}, we computed the 2–10~keV model flux for reference. The \texttt{fakeit} command then folds our current \textsc{XSPEC} model, including relativistic reflection components through the RMF and ARF, adds Poisson noise, and samples the background to produce a realistic \texttt{PHA} file. For spectral fitting, we adopted the chi statistic and ignore channels below 1.5~keV and above 75~keV, where the \textit{NuSTAR} response is either poorly calibrated or background‐dominated.  The resulting mock spectrum closely mimics a high-quality \textit{NuSTAR} observation and provides the basis for parameter estimation and residual analysis.

The simulated spectrum and data–model comparison are shown in Fig.~\ref{mock_spectrum_lambda00}. The mock dataset described above was produced from a Kerr-like wormhole reflection template evaluated at $\lambda=0$ (equivalent to a rapidly rotating Kerr black hole with $a_*=0.998$) under \textit{NuSTAR} instrumental conditions. Fitting this dataset with our convolution model \texttt{kwconv*xillverCp} yields an excellent description of the folded spectrum. The best-fit model reproduces the broad continuum and iron-band features with only small, statistically acceptable residuals. In particular, the fit recovers the input parameter values of the simulation to within the formal statistical uncertainties returned by the fit, demonstrating that our model parametrization is fully compatible with the Kerr limit of our templates. 

{To validate the consistency of our customized \texttt{kwline}\footnote{https://github.com/bhXray/kwline} implementation, we performed a cross-check by fitting a standard Kerr mock spectrum using our wormhole convolution model (\texttt{kwconv}). As shown in Table~\ref{tab:validation_kerr}, when the deformation parameter $\lambda$ is allowed to vary, the best-fit values return to the Kerr {black hole} limit ($\lambda \approx 0$) with high precision. Specifically, for a mock \textit{NuSTAR}-like spectrum generated with $a_* = 0.998$ and $i=75^\circ$, our model achieved a reduced chi-squared of $\chi^2_{\rm red} \approx 1.06$, with all physical parameters (spin, inclination, and emissivity index) being recovered within their $1\sigma$ confidence intervals. This successful recovery demonstrates that our radial integral corrected analytic framework is perfectly backward-compatible with the Kerr metric.
To further scrutinize the "metric mimicry" effect, we fitted the same mock data using the standard modular Kerr model (\texttt{kerrconv * xillverCp}), which result is shown in Fig.~\ref{fig:mock_kerrfitkerr} and Table~\ref{tab:validation_kerr}. This comparison revealed that the standard Kerr metric also provides an excellent fit ($\chi^2_{\rm red} \approx 1.037$) and accurately recovers the input parameters. However, the \texttt{kerrconv} fitting results present a distinct geometric scenario compared to the \texttt{kwconv} model. The Kerr model characterizes the source as having a very extended disk structure, with the radial extent of the disk stretching from the ISCO to an outer radius of approximately $400~r_g$.This extended disk interpretation, coupled with a steep inner emissivity index and a highly unconstrained break radius, suggests that the Kerr model mimics the specific redshift distribution of the wormhole metric by redistributing the emission across a wider radial range. }

\begin{figure}
    \centering
    \includegraphics[width=0.95\linewidth]{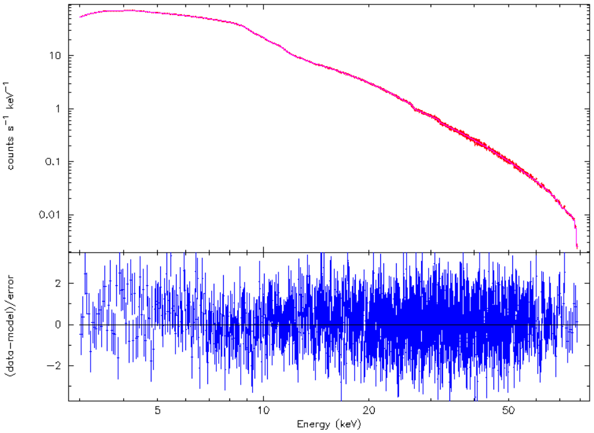}
    \caption{Mock \textit{NuSTAR} spectrum of bright X-ray binary generated from our Kerr-like wormhole theoretical reflection spectrum with throat parameter $\lambda=0$ (which corresponds to a rapidly rotating Kerr black hole with $a_*=0.998$), folded through realistic \textit{NuSTAR} RMF and ARF and background files, and fitted with our wormhole convolution model \texttt{kwconv*xillverCp}. {Top:} Folded spectrum with the best-fit \texttt{kwconv*xillverCp} (Kerr-like wormhole model) model overplotted. {Bottom:} Data-to-model residuals expressed in units of the data error, $(\mathrm{data}-\mathrm{model})/\mathrm{error}$. 
    }
    \label{mock_spectrum_lambda00}
\end{figure}

\begin{figure}
    \centering
    \includegraphics[width=0.95\linewidth]{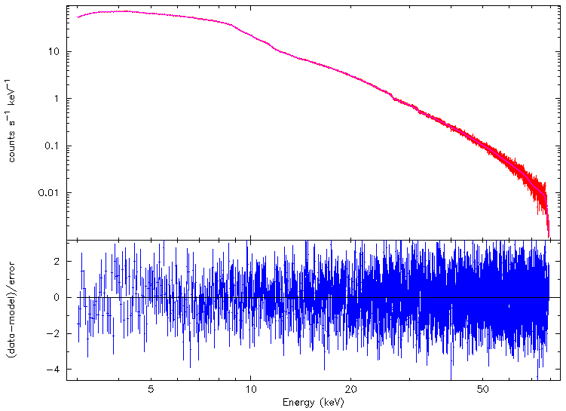}
    \caption{{Same as Fig.~\ref{mock_spectrum_lambda00} but fitted with the Kerr black hole covolution model \texttt{kerrconv*xillverCp}.}}
    \label{fig:mock_kerrfitkerr}
\end{figure}

\subsection{Exploratory wormhole versus\ black hole via synthetic spectrum}
\label{ssec:wormhole}

As an initial study into distinguishing Kerr-like wormholes from standard Kerr black holes, we simulated a long FPMA exposure using the wormhole iron line convolved X-ray spectrum computed in Sect.~\ref{sec:convolved_spectrum}. Although a full reflection-spectrum analysis with real data lies beyond the scope of this work, our exploratory test using a synthetic spectrum assesses whether current facilities could discriminate between the two geometries. All simulations and spectral fits are performed in \textsc{XSPEC} using the standard \textit{NuSTAR} RMF and ARF and background files described above. We did not target a particular source. Instead, we adopted representative parameters for a high-spin Kerr-like wormhole source: a thermal Comptonization continuum with photon index $\Gamma=2$ and an electron temperature $kT_e = 300$ keV. To define the source's intensity, we calculated the model flux in the $2-10$ keV energy range for an exposure time of $50\,$ks, which serves as a baseline for the \texttt{fakeit} procedure. The simulated spectra are rebinned to a minimum of 20 counts per bin and fitted using $\chi^2$ minimization over 0.7--10~keV. Table~\ref{tab:combined-fitting-results} summarizes the best-fit parameter values for the model families considered. Unless stated, we allowed the continuum photon index, normalizations, iron line normalization, spin parameter, and disk inclination to vary, while redshifts was fixed.

To ensure a robust comparison, we explored the parameter space using the standard Levenberg-Marquardt $\chi^2$ minimization algorithm within \textsc{XSPEC}. We allowed key physical parameters including the spin values $a_* \in [0, 0.9999]$, inclination angles $i \in [5^\circ, 89^\circ]$, emissivity indices ($q_{\rm in}, q_{\rm out}$), disk radii ($r_{\rm in}, r_{\rm out}$), ionization ($\log \xi$), iron abundance $A_{\rm Fe} \in [0.05, 5]$ in the unit of the solar abundance, and the reflection fraction to vary freely across their full physically plausible ranges.

A key finding from our simulations is that both the wormhole convolution model (\texttt{kwconv*xillverCp}) and the standard Kerr black hole convolution model (\texttt{kerrconv*xillverCp}) provide excellent fits to the mock data. Model 1 (\texttt{kwconv*xillverCp}) yields a reduced $\chi^2 \approx 1.02$ with a null-hypothesis probability of $0.28$, successfully recovering the input deformation $\lambda \approx 0.9$. Strikingly, Model 2 (\texttt{kerrconv*xillverCp}) achieves a nearly identical fit quality with $\chi^2_{\text{red}} \approx 1.04$ and a null-hypothesis probability of $0.13$, even though it lacks the parameter $\lambda$. The residuals for both models (as seen in Figs.~\ref{mock_spectrum_lambda09} and \ref{mock_spectrum_fitKerr}) remain largely flat across the 3–79 keV band. This suggests that at the current spectral resolution of \textit{NuSTAR}, Kerr-like wormholes can effectively mimic Kerr black holes, making them observationally indistinguishable through simple convolutional reflection modeling.

To further investigate whether a Kerr black hole-based reflection description can reproduce the mock data developed from a Kerr-like wormhole, we also fited the same simulated spectrum with the Kerr black hole-based relativistic reflection model \texttt{relxill} (\texttt{tbabs*relxillCp}) \citep{Dauser:2010ne, Dauser:2022zwc}. The \texttt{relxill} framework couples the angle-dependent \texttt{xillver} reflection calculations with a relativistic ray-tracing kernel (the \texttt{relline} family) so that, for each disk element, the appropriate angle-resolved local reflection spectrum is selected.
{Specifically, we employed the \texttt{relxillCp} variant, which differs from the standard \texttt{relxill} model by assuming a thermally Comptonized continuum (\texttt{nthComp}) as the primary ionizing source rather than a simple power law with an exponential cutoff. Given that our mock wormhole spectrum was generated using a Comptonization primary (\texttt{xillverCp}), choosing \texttt{relxillCp} is the most appropriate approach to ensure that any resulting fit residuals arise solely from differences in the relativistic blurring and space-time geometry, rather than a mismatch in the assumed spectral shape of the corona.}

\begin{figure}
    \centering
    \includegraphics[width=1.0\linewidth]{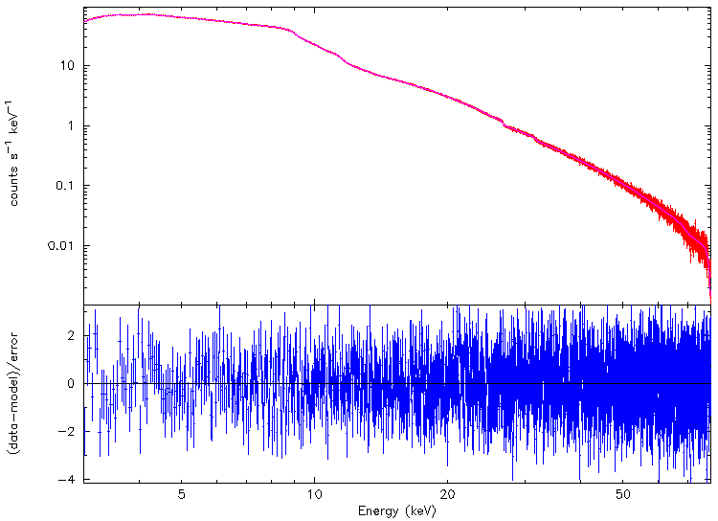}
    \caption{
    Mock \textit{NuSTAR} spectrum for a bright AGN or compact source generated from our Kerr-like wormhole reflection template with throat parameter $\lambda=0.9$ and spin $a_*=0.998$, folded through realistic \textit{NuSTAR} RMF and ARF and background files, and fitted with the convolution model \texttt{kwconv*xillverCp} (Kerr-like wormhole model). {Top:} Folded spectrum with the best-fit \texttt{kwconv*xillverCp} model overplotted. {Bottom:} Data-to-model residuals in units of the data error, $(\mathrm{data}-\mathrm{model})/\mathrm{error}$.
    }
    \label{mock_spectrum_lambda09}
\end{figure}

\begin{figure}
    \centering
    \includegraphics[width=1.0\linewidth]{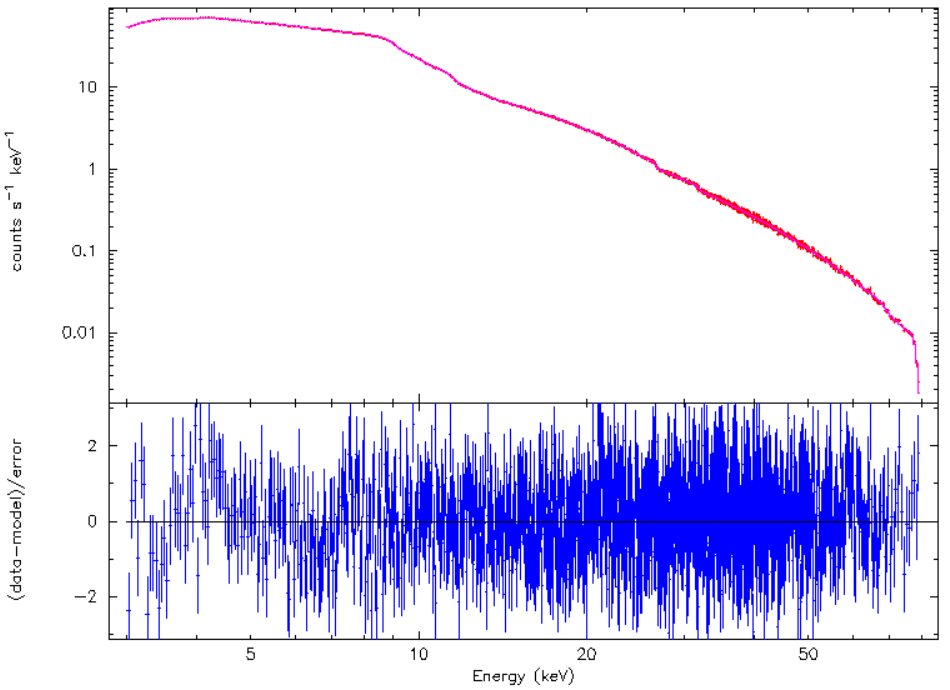}
    \caption{Same as Fig.~\ref{mock_spectrum_lambda09} but fitted with the Kerr black hole covolution model \texttt{kerrconv*xillverCp}.
    }
    \label{mock_spectrum_fitKerr}
\end{figure}

For this test, we deliberately omited the background-exposure term to examine the intrinsic performance of the model under the same instrumental responses and binning used above. The fitting result is shown in Fig.~\ref{mock_spectrum_fitrel}. 
The detailed numerical outcome of forcing the Kerr-based \texttt{relxillCp} model onto the wormhole mock data is presented in Table~\ref{tab:combined-fitting-results}.
The fit returns $\chi^{2} = 2803.18$ for 1931 bins (with 1916 degrees of freedom), corresponding to a reduced $\chi^2_{\rm red} \simeq 1.46$. Most importantly, the null-hypothesis probability is extremely low ($1.28 \times 10^{-36}$), indicating a formal statistical failure of the model.
The fitting result seen in Table~\ref{tab:combined-fitting-results} displays anomalous values of several parameters. In particular, the inner emissivity index {reached} its maximum allowed value ($q_\text{in} = 10.0$), while the reflection fraction {had} a high value of \textit{refl\_frac} $\simeq 10.0$. Some parameters, such as the hydrogen column density ($n_H$) and the inner radius ($R_{in}$), returned unconstrained or unphysical values (reported with errors of $-1.0$). These behaviors are features of a model attempting to compensate for a fundamental mismatch in the intrinsic spectral morphology. Quantitatively, the \texttt{relxillCp} fit is significantly worse than the convolutional models (\texttt{kwconv} and \texttt{kerrconv}) and yields strong structured residuals concentrated in the $5-10$~keV iron line complex and the $10-20$~keV Compton hump region. Despite allowing key physical parameters including the emissivity indices, disk radii, ionization ($\log\xi \simeq 3.19$), and iron abundance ($A_{\rm Fe} \simeq 1.49$) to vary freely, no combination of parameters within the standard Kerr {black hole} framework could resolve these localized residuals. The failure of the \texttt{relxillCp} model, compared to the relative success of the convolutional \texttt{kerrconv} model, provides strong evidence that the modeling framework itself (e.g., the treatment of angle-dependent reflection and the reflection fraction) plays a crucial role. The anomalous best-fit values should be interpreted as signs that the standard integrated \texttt{relxillCp} model is unable to mimic the specific iron line and reflection morphology induced by the wormhole geometry under these simulated conditions.

{The significant discrepancy in {fitting} quality between the two Kerr-based descriptions, the convolutional \texttt{kerrconv*xillverCp} (Model 2) and the integrated \texttt{relxillCp} (Model 3), yields a more nuanced conclusion. While \texttt{kerrconv*xillverCp} provides a statistically acceptable fit ($\chi^2_{\rm red} \simeq 1.04$), \texttt{relxillCp} fails significantly ($\chi^2_{\rm red} \simeq 1.46$) with structured residuals.
We therefore interpreted these results as an indication of the sensitivity of the modeling framework rather than a definitive exclusion of the Kerr geometry. The \texttt{relxill} framework is fundamentally more rigid; it enforces a specific, self-consistent coupling between the ray-tracing kernel and the angle-dependent reflection tables. In contrast, the convolutional approach (used both for generating the mock data and in Model 2) treats the relativistic blurring as a post-processing step applied to an angle-averaged reflection spectrum.
The failure of \texttt{relxillCp} to reproduce the mock data likely stems from this architectural difference. Specifically, the wormhole-induced line morphology, when processed through a simple convolution, may lack certain angle-resolved features that \texttt{relxillCp} expects to see in a "pure" Kerr {black hole} reflection signal. Consequently, the model attempts to compensate for this mismatch by invoking extreme parameter values, such as a high emissivity index ($q_{\rm in}=10.0$) and a high reflection fraction ($refl\_frac \simeq 10.0$).
These results suggest that while existing convolutional models for Kerr-like wormholes and Kerr black holes exhibit high observational degeneracy at \textit{NuSTAR} resolutions, more integrated models like \texttt{relxill} are more sensitive to the underlying modeling assumptions. The anomalous best-fit values in Table~\ref{tab:combined-fitting-results} should therefore be viewed as a signature of modeling paradigm mismatch where a self-consistent Kerr framework struggles to fit a simplified convolutional wormhole template, rather than a direct physical measurement of the space-time's properties.}

Overall, our analysis suggests that while Kerr-like wormholes and Kerr black holes may appear degenerate under traditional convolutional modeling, they can potentially be discriminated through more rigid, self-consistent reflection frameworks. Such distinctions require not only high-quality data but also a careful choice of the underlying modeling assumptions to move beyond simple post-processing approximations.

\begin{figure}
    \centering
    \includegraphics[width=1.0\linewidth]{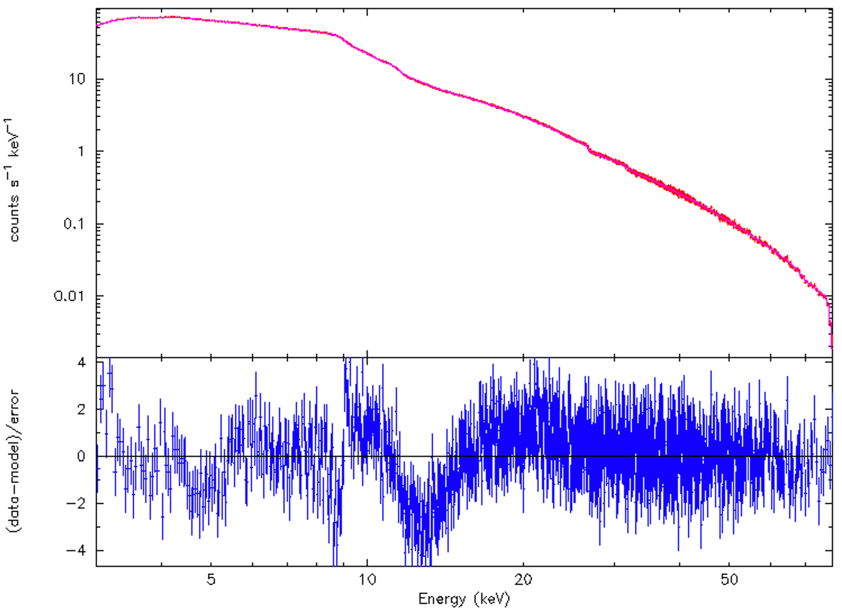}
    \caption{Same as Fig.~\ref{mock_spectrum_lambda09} but fitted with the Kerr black hole-based relativistic reflection model \texttt{relxillCp}.
    }
    \label{mock_spectrum_fitrel}
\end{figure}

\section{Summary and discussion}
\label{sec:Summary}

\renewcommand{\theequation}{5.\arabic{equation}} \setcounter{equation}{0}

In this work, we developed a relativistic reflection framework that explicitly incorporates Kerr-like traversable wormhole metrics into standard X-ray reflection modeling by implementing two \textsc{XSPEC} modules (\texttt{kwline} for $\delta$-function iron lines and \texttt{kwconv} for convolving full rest-frame reflection spectra). Using a fine grid over the spin and throat radius, we produced libraries of ray-traced line profiles and convolved reflection spectra, then folded these through realistic \textit{NuSTAR} RMF and ARF and background files to create mock observations for spectral fitting. This pipeline enables a direct, end-to-end comparison between wormhole and Kerr black hole predictions under identical instrumental and analysis assumptions.

Our ray-traced calculations show that the throat parameter $\lambda$ reveals a complex, non-monotonic relationship with the line width. Specifically, as the space-time deviates from the Kerr metric ($\lambda=0$), the red wing progressively narrows due to the retreating inner disk edge ($r_{in} \approx 3.32M$ for $\lambda=0.9$), which effectively removes the most highly redshifted emission from the deepest regions of the potential well. This produces the characteristic narrower and less redshifted iron feature observed in our wormhole cases compared to standard Kerr black holes at near-extremal spins.
We find that the ability to distinguish these features via observations is critically dependent on the disk's physical properties, particularly the emissivity index $q$. At steep emissivity gradients ($q=6$), the radiation is concentrated in the innermost regions, which maximizes the spectral discrepancy between the extended red wing of a Kerr black hole and the truncated profile of a wormhole. Conversely, in the $q=0$ limit where emission is dominated by the outer disk, these relativistic effects are substantially diluted, which renders the two geometries nearly indistinguishable. 
Furthermore, the minimum width critical point identified at $\lambda \approx 0.6$ demonstrates that iron line spectroscopy is sensitive to a specific window of deformation where relativistic effects, such as gravitational redshift and Doppler boosting, are most distinctively imprinted. 

Our results demonstrate that Kerr-like wormholes imprint characteristic modifications on the Fe K$\alpha$ profile, typically producing narrower profiles with a suppression of the red wing compared to Kerr black holes at fixed parameters. These effects result from the throat's alteration of the image-plane mapping and the redshift distribution of photons. However, a key finding of our simulated spectral fitting is the high degree of observational degeneracy between convolutional models. At the current spectral resolution of \textit{NuSTAR}, we find that a standard Kerr black hole convolution model (\texttt{kerrconv*xillverCp}) can provide a statistically acceptable fit ($\chi^2_{\rm red} \simeq 1.04$, $P_{\rm null} = 0.13$) to a mock wormhole spectrum with a large throat parameter ($\lambda=0.9$). This suggests that simple convolutional templates may not be sufficient to uniquely distinguish these geometries in practical observational scenarios.

To further assess model discrimination, we tested a more physically integrated relativistic reflection model, \texttt{relxillCp}. Unlike the convolutional models, the \texttt{relxillCp} fit to the mock wormhole data resulted in a significant statistical failure ($\chi^2 \simeq 2803.18$ for 1916 d.o.f., $P_{\rm null} \simeq 10^{-36}$) and left structured residuals in the 5--20~keV band. We interpreted this as a sign of a modeling paradigm mismatch: the integrated \texttt{relxill} framework enforces a self-consistent coupling between the ray-tracing kernel and angle-dependent reflection that is far more sensitive to the intrinsic line morphology. While the convolutional models exhibit degeneracy due to their simpler mathematical structure, the failure of the self-consistent Kerr black hole model to mimic the wormhole signal implies that fundamental geometric differences are, in principle, detectable given sufficiently rigorous modeling.

We performed robustness checks that identify important caveats. The success of Model 2 (\texttt{kerrconv*xillverCp}) compared to the failure of Model 3 (\texttt{relxillCp}) highlights that model selection is just as critical as the underlying space-time geometry. The anomalous parameter values recovered in the \texttt{relxillCp} fit, such as an emissivity index, $q_{\rm in}$ {, reached} at 10.0 with a high reflection fraction, should be read as indicators of the model's struggle to adapt to the wormhole-induced iron line morphology rather than as physically meaningful measurements. Future archival searches must therefore (i) employ full error-region mapping (MCMC or \texttt{steppar}), (ii) carefully evaluate the sensitivity to different modeling frameworks (convolutional vs. integrated), and (iii) perform cross-instrument fits (e.g., with \textit{XRISM}) to resolve these subtle features.

In summary, while current X-ray facilities can face challenges in resolving the degeneracy between Kerr-like wormholes and black holes through standard convolutional templates, the use of more integrated, self-consistent reflection models reveals significant morphological differences. The wormhole relativistic transfer functions implemented here produce clear modifications to the Fe K$\alpha$ region that can be distinguished if the modeling framework is sufficiently sensitive to the underlying physical self-consistency. We therefore advocate for the inclusion of horizonless template families in standard reflection analyses as a vital test of strong-field gravity in the era of high-precision X-ray missions.

\begin{acknowledgements}

We would like to thank Dr. Andrew Mummery, for a constructive and very detailed instruction. {We appreciate the constructive comments provided by the anonymous reviewers, which have improved the manuscript.}
This research is supported by the National Key R\&D Program of China (Grant No.\,2023YFE0101200), the National Natural Science Foundation of China (Grant No.\,12273022, 12192220, 12133008), and the Shanghai Municipality Orientation Program of Basic Research for International Scientists (Grant No.\,22JC1410600).
T.Z. is supported by the National Natural Science Foundation of China under Grants No.\,12275238, the Zhejiang Provincial Natural Science Foundation of China under Grants No.\,LR21A050001 and No.\,LY20A050002, the National Key R\&D Program of China under Grant No.\,2020YFC2201503, and the Fundamental Research Funds for the Provincial Universities of Zhejiang in China under Grant No.\,RF-A2019015. 
\end{acknowledgements}

\bibliographystyle{aa}
\bibliography{Refs}

\newpage

\begin{appendix}

\section{Brief introduction of the properties of Kerr-like wormhole}\label{sec:KerrlikeWH}
\numberwithin{equation}{section}

In this section we review the basic ingredients needed for the construction of accretion disk images in wormhole geometries. We first specify the space-time metrics, and then record some known analytical results regarding accretion disks in these backgrounds. This section has review material, meant to set up the scenarios to be used later in this paper.

\subsection{Kerr-like wormhole}

The line element of the wormhole was obtained in \cite{Bueno:2017hyj}, and is given by
\bqn
\begin{aligned}
ds^2 = &-\left(1 - \frac{2Mr}{\Sigma}\right) dt^2 - \frac{4Mar\sin^2\theta}{\Sigma} dtd\phi + \frac{\Sigma}{\hat{\Delta}} dr^2 + \Sigma d\theta^2 \\
&+ \left(r^2 + a^2 + \frac{2Ma^2r\sin^2\theta}{\Sigma}\right)\sin^2\theta d\phi^2, \\
\Sigma \equiv &\ r^2 + a^2\cos^2\theta, \quad \hat{\Delta} \equiv r^2 - 2M(1 + {\lambda}^2)r + a^2.
\label{Eq:metric}
\end{aligned}
\eqn
The parameters \(a\) and \(M\) are related, respectively, to the spin and mass of the wormhole, and \(\lambda\) is the deformation parameter whose value signifies the deviation of the metric of Eq.~\eqref{Eq:metric} from the Kerr black hole. For \(\lambda = 0\), we recovered the Kerr black hole and the corresponding black hole horizons are given by \(\hat{\Delta}(r_{\pm}) = 0\), i.e., by \(r_{\pm} = M \pm \sqrt{M^2 - a^2}\). For \(\lambda \neq 0\), however, the space-time represents a wormhole and the wormhole throat \(r_{\rm th}\) is given by \(\hat{\Delta}(r_{\rm th}) = 0\), i.e., by
\bqn
r_{\rm th} = M(1 + \lambda^2) + \sqrt{M^2(1 + \lambda^2)^2 - a^2}. \label{rth}
\eqn

Note that the Eq.~\eqref{rth} necessarily implies that \(r_{\rm th} > r_{+}\), where \(r_{+}\) is the outer horizon of the Kerr black hole. One can easily verify that when all the throat effects of the wormhole are absent ($\lambda=0$), the metric of the Kerr-like wormhole goes back to the Kerr space-time exactly.

{The geometric difference between the Kerr black hole and the Kerr-like wormhole is illustrated through their embedding surfaces in Fig.~\ref{diagram_wormhole}. While the Kerr metric (blue) possesses an event horizon, the wormhole metric (orange) represents a traversable bridge connecting two symmetric regions of space-time via the throat $r_{\rm th}$. This structural deviation is fundamentally governed by the deformation parameter $\lambda$.
However, for asymmetric wormhole, the space-times on the two sides are different. In this work, we do not consider this latter case for simplicity.}

\begin{figure}
    \centering
    \includegraphics[width=0.9\linewidth]{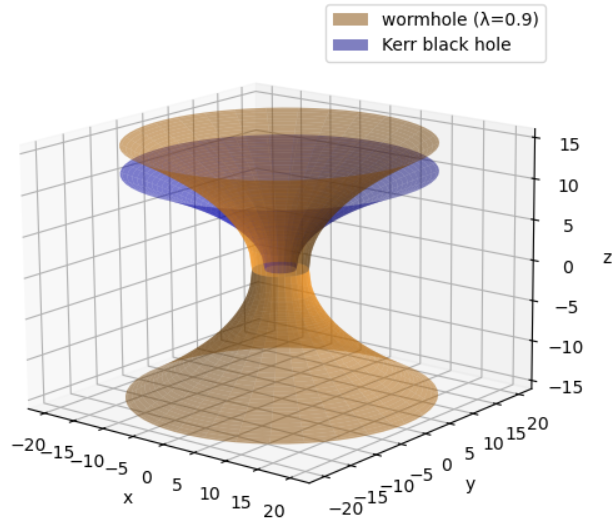}
    {\caption{Embedding surfaces of the equatorial slice for a Kerr black hole (blue) and a Kerr-like wormhole (orange, $\lambda$=0.9).}}
    \label{diagram_wormhole}
\end{figure}

\subsection{Orbital motion and ISCOs} 
\label{subsec:ISCO}

The most relevant component of the orbit for the study of accretion flows is the angular momentum of a circulating fluid, since this is an excellent approximation of the fluid dynamics in the stable disk regions.

The accretion disk is formed by particles moving in circular orbits around a compact object, whose physical properties and electromagnetic radiation characteristics are determined by the space-time geometry around the central compact object. For the purpose of studying the emission properties of accreting gas, let us first consider the evolution of a massive particle in the Kerr-like wormhole space-time. We start with the Lagrangian of the particle,
\bqn
\mathscr{L} = \frac{1}{2}g_{\mu \nu} \frac{d x^\mu} {d \tau } \frac{d x^\nu}{d \tau},
\eqn
where $\tau$ denotes the affine parameter of the world line of the particle. For a massless particle, we have $\mathscr{L}=0$, and for a massive one, $\mathscr{L} <0$. Then, the generalized momentum $p_\mu$ of the particle can be obtained via
\bqn
p_{\mu} = \frac{\partial \mathscr{L}}{\partial \dot x^{\mu}} = g_{\mu\nu} \dot x^\nu,
\eqn
which leads to four equations of motion for a particle with energy $\tilde{E}$ and angular momentum $\tilde{L}$,
\bqn
p_t &=& g_{tt} \dot t +g_{t\phi} \dot \phi  = - \tilde{E},\\
p_\phi &=&g_{\phi t} \dot t+ g_{\phi \phi} \dot \phi = \tilde{L}, \\
p_r &=& g_{rr} \dot r,\\
p_\theta &=& g_{\theta \theta} \dot \theta.
\eqn
Here, an overdot denotes the derivative with respect to the affine parameter $\lambda$ of the geodesics. From these expressions, we obtain 
\bqn
\dot t =  \frac{g_{\phi\phi} \tilde{E}+g_{t\phi}\tilde{L}  }{ g_{t\phi}g_{\phi t}- g_{tt}g_{\phi\phi} } ~\\
\dot \phi = \frac{\tilde{E}g_{t\phi}+ g_{tt} \tilde{L}}{g_{tt}g_{\phi\phi}-g_{t\phi}g_{\phi t}}.
\eqn
For the conservation of the rest-mass, we have $ g_{\mu \nu} \dot x^\mu \dot x^\nu = -1$. Substituting $\dot t$ and $\dot \phi$ we can get
\bqn
g_{rr} \dot r^2 + g_{\theta \theta} \dot \theta^2 = -1 - g_{tt} \dot t^2  - g_{\phi\phi}\dot \phi^2 -2g_{t \phi}\dot{t}\dot{\phi}.~~
\eqn
{To investigate the gravitational properties of the Kerr-like wormhole, we consider the motion of a test particle in the equatorial plane ($\theta = \pi/2$). The radial equation of motion can be derived from the Lagrangian and the normalization condition of the four-velocity, expressed as:
\bqn
    g_{rr} \dot{r}^2 = V_{\rm eff}(r, E, L),
\eqn
where $E$ and $L$ are the conserved specific energy and angular momentum per unit rest mass, respectively. A crucial property of this metric is that the radial metric coefficient $g_{rr}$ acts only as a scaling factor for the radial velocity $\dot{r}$ and does not enter the expression for the effective potential $V_{\rm eff}$. Specifically, for any stationary and axisymmetric space-time, the effective potential is determined solely by the $g_{tt}, g_{t\phi}$, and $g_{\phi\phi}$ components:
\bqn
    V_{\rm eff}(r, E, L) = \frac{E^2 g_{\phi\phi} + 2ELg_{t\phi} + L^2 g_{tt}}{g_{t\phi}^2 - g_{tt}g_{\phi\phi}} - 1.
\eqn}

{For the Kerr-like wormhole considered here, these three metric components are identical to those of the standard Kerr metric. Consequently, the conditions for stable circular orbits, defined by $V_{\rm eff}(r) = 0$ and $V_{\rm eff}'(r) = 0$, yield solutions for $E(r)$ and $L(r)$ that are indistinguishable from those in the Kerr black hole case. This property implies an ``observational mimicry'' regarding the dynamics of the accretion disk.}

\begin{figure}
    \centering
    \includegraphics[width=1.0\linewidth]{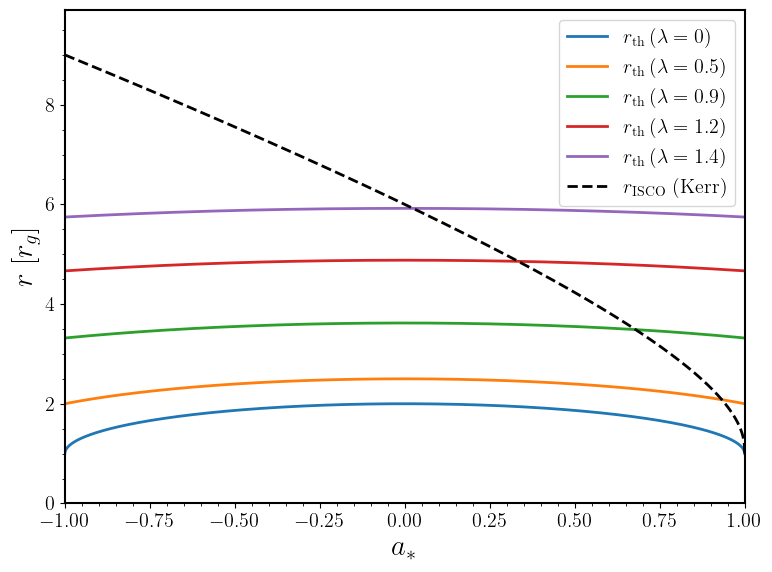}
    \caption{Comparison of the ISCO and throat size of the Kerr-like wormhole for the different $\lambda$ values.}
    \label{ISCO_rth}
\end{figure}

The ISCO marks the inner edge of a Keplerian accretion disk and is determined by the additional condition $V_{\rm eff}''(r) = 0$. Since $V_{\rm eff}$ is independent of the modification of $g_{rr}$ (where the modified function $\hat{\Delta}$ replaces the standard Kerr $\Delta$), the ISCO radius $r_{\rm ISCO}$ in our wormhole space-time remains the same as in the Kerr case for a given mass $M$ and spin $a$. Following \citep{Bardeen:1972fi}, the analytical expression for the equatorial ISCO radius is:
\bqn
    r_{\rm ISCO} = M \bigl\{ 3 + Z_2 \mp \sqrt{(3 - Z_1)(3 + Z_1 + 2Z_2)} \bigr\},
\eqn
where the auxiliary quantities $Z_1$ and $Z_2$ are defined by the dimensionless spin $a_* = a/M$:
\bqn
    \begin{aligned}
        Z_1 &= 1 + \left(1 - a_*^2\right)^{1/3} \left[(1 + a_*)^{1/3} + (1 - a_*)^{1/3}\right], \\
        Z_2 &= \sqrt{3a_*^2 + Z_1^2}.
    \end{aligned}
\eqn
The negative (positive) sign denotes prograde (retrograde) orbits.

{Another critical observable is the orbital angular velocity $\Omega \equiv d\phi/dt$, which represents the frequency of the particle as seen by a distant observer. For a circular orbit, this is given by:}
\bqn
    \Omega = \frac{-\partial_r g_{t\phi} \pm \sqrt{(\partial_r g_{t\phi})^2 - (\partial_r g_{tt})(\partial_r g_{\phi\phi})}}{\partial_r g_{\phi\phi}}.
\eqn
Consistent with our previous discussion, since the modification in the radial sector does not appear in these metric components or their radial derivatives, the angular velocity simplifies to the classic Keplerian frequency:
\bqn
    \Omega = \frac{M^{1/2}}{r^{3/2} \pm a M^{1/2}}.
\eqn
{This result highlights that the wormhole parameter $\lambda$ is purely geometric and confined to the radial sector; it does not alter the local dynamics or frequencies of circular motion in the equatorial plane.}

{For the circular orbits discussed above to be physically accessible, the ISCO radius must lie outside or coincide with the wormhole throat, satisfying $r_{\rm ISCO} \ge r_{\rm th}$. Figure~\ref{ISCO_rth} provides a direct comparison between the ISCO radius and the throat radius $r_{\rm th}$ as a function of the spin parameter $a_*$ for various $\lambda$ values. In this context, positive values of $a_*$ denote prograde orbits, where the angular momentum of the accretion flow is aligned with the spin of the central object, while negative values represent retrograde orbits. Considering the dynamical stability and the typical evolution of astrophysical accretion systems, our subsequent analysis focuses exclusively on prograde configurations ($a_* \ge 0$). The critical condition $r_{\rm th} = r_{\rm ISCO}$ defines the maximum allowable value for the wormhole parameter, denoted as $\lambda_{\rm max}$. By solving this equality, we obtain an explicit analytic relation for the upper bound of the wormhole parameter:
\bqn
    \lambda_{\rm max}(a_*) = \sqrt{ \frac{r_{\rm ISCO}^2 + a_*^2}{2\,r_{\rm ISCO}} - 1 },
\eqn
where $r_{\rm ISCO}$ is the standard Kerr ISCO radius. As illustrated in Fig.~\ref{para_distribution}, this relation divides the parameter space into two physically distinct regimes regarding the inner edge of the accretion disk.}

In the first regime, where $\lambda \le \lambda_{\rm max}$, the ISCO is located in the outer region of the space-time ($r_{\rm ISCO} \ge r_{\rm th}$). Under these conditions, the accretion disk structure is identical to that of a Kerr black hole, and its inner edge is conventionally determined by the marginal stability condition $V_{\rm eff}''(r) = 0$. In this case, the wormhole behaves as an effective black hole mimicker for several orbital observables and the iron line profiles.

{Conversely, in the regime where $\lambda > \lambda_{\rm max}$, the formal ISCO radius would lie below the throat. As discussed by \citet{Paul:2019trt}, the circular orbit at the throat itself then acts as the marginal stable orbit. This transition occurs because the inverse radial metric component $g_{rr}^{-1}$ vanishes at the throat $r = r_{\rm th}$, which ensures that the conditions for a circular orbit, $V_{\rm eff}=0$ and $V_{\rm eff}'=0$, automatically fulfill the requirement for a stable trajectory where $\ddot{r}=0$. Consequently, for sufficiently large $\lambda$, the accretion disk is no longer truncated by orbital instability but is instead limited by the topological boundary of the wormhole throat. This provides a potential observational signature to distinguish such wormholes from standard Kerr black holes.}

\begin{figure}
    \centering
    \includegraphics[width=0.8\linewidth]{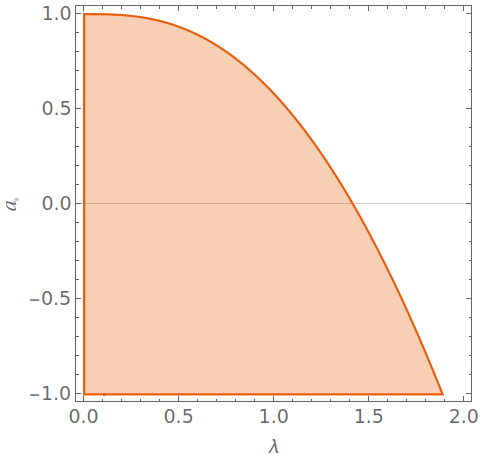}
    \caption{Parameter distribution of the Kerr-like wormhole as a function of dimensionless spin parameter $a_*$ for the different $\lambda$ values.}
    \label{para_distribution}
\end{figure}

{
\subsection{Null geodesics and ray-tracing algorithm}
\label{subsec:nullgeodesics}
To rigorously treat photon propagation from the accretion disk to a distant observer, we employed a semi-analytic ray-tracing technique. The Kerr-like wormhole metric (Eq. \eqref{Eq:metric}) possesses a Killing tensor, ensuring that the Hamilton-Jacobi equation is separable. This allows us to express the null geodesics as a set of first-order decoupled equations of motion. Following the frameworks of \citet{Amir:2018pcu}, the equations for a photon with energy $E$, axial angular momentum $L_z$, and Carter constant $\mathcal{Q}$ are given by
\begin{align}
    \Sigma \frac{dt}{d\sigma} &= -a(aE\sin^2\theta - L_z) + \frac{(r^2+a^2)[E(r^2+a^2)-aL_z]}{\Delta}, \\
    \Sigma \frac{dr}{d\sigma} &= \pm \sqrt{\mathcal{R}(r)}, \\
    \Sigma \frac{d\theta}{d\sigma} &= \pm \sqrt{\Theta(\theta)}, \\
    \Sigma \frac{d\phi}{d\sigma} &= -\left( aE - \frac{L_z}{\sin^2\theta} \right) + \frac{a[E(r^2+a^2)-aL_z]}{\Delta},
\end{align}
where $\sigma$ is the affine parameter and $\Sigma = r^2 + a^2 \cos^2\theta$. Crucially, in our wormhole geometry, the radial and angular potentials take the following forms:
\begin{align}
    \mathcal{R}(r) &= \frac{\hat{\Delta}}{\Delta} \left\{ [E(r^2+a^2) - aL_z]^2 - \Delta[\mathcal{Q} + (L_z - aE)^2] \right\}, \\
    \Theta(\theta) &= \mathcal{Q} + \cos^2\theta \left( a^2E^2 - \frac{L_z^2}{\sin^2\theta} \right).
\end{align}
In these expressions, $\Delta = r^2 - 2Mr + a^2$ is the standard Kerr horizon function, which appears in the $t$ and $\phi$ sectors of the metric. The wormhole modification enters exclusively through the function $\hat{\Delta} = r^2 - 2M(1+\lambda^2)r + a^2$ in the radial potential $\mathcal{R}(r)$. As the referee correctly noted, the modification of $g_{rr}$ does not influence the effective potential for circular orbits in the disk (see Sec. \ref{subsec:ISCO}); however, the ratio $\hat{\Delta}/\Delta$ acts as a critical scaling factor for the radial null geodesics. This ensures that while the accretion disk's dynamics (ISCO and orbital frequency) remain Kerr-like, the propagation of light is sensitive to the wormhole throat parameter $\lambda$, leading to the observed differences in the iron line profiles.
}

\begin{figure}
    \centering
    \includegraphics[width=0.85\linewidth]{Figures/trajectory.png}
    \caption{{Photon trajectories in the Kerr-like wormhole space-time with $a = 0.998, \lambda=0.9$ for an observer at $\theta_0=80^\circ$. Each trajectory is uniquely determined by its initial coordinates $(\alpha, \beta)$ on the observer's celestial plane. The trajectories contains rays that are deflected around the object as well as rays that approach (and may traverse) the throat, demonstrating the nontrivial light propagation in the wormhole geometry. The resulting trajectories are shown in 3D Cartesian coordinates in units of $GM/c^2$.}}
    \label{trajectory}
\end{figure}

{Our numerical implementation employs a customized version of the \textsc{YNOGK} code \citep{Yang:2013yoa}, which is built upon \textsc{GEOKERR} \citep{Dexter:2009fg}. To maintain the high computational efficiency of the original analytic framework while ensuring physical accuracy in the wormhole geometry, we employed a semi-analytical path-integration scheme.}

Specifically, the modification in the radial metric, which is characterized by the deformation factor $v(r) \equiv \sqrt{\Delta/\hat{\Delta}}$ is accounted for by decomposing the geodesic integrals into a standard Kerr analytical part and a wormhole-induced residual part. For any radial integral $I$ (e.g., for coordinate time $t$, azimuth $\phi$, or affine parameter $\lambda$), the integration is expressed as
\bqn
\mathcal{I} = \int^{r} \mathcal{K}(r) \sqrt{\frac{\Delta}{\hat{\Delta}}} dr = \int^{r} \mathcal{K}(r) dr + \int^{r} \mathcal{K}(r) \left( \sqrt{\frac{\Delta}{\hat{\Delta}}} - 1 \right) dr, \label{eq:decomp} \nb\\
\eqn
where $\mathcal{K}(r)$ denotes the standard Kerr radial kernel (e.g., $1/\sqrt{\mathcal{R}}$).The first term on the right-hand side of Eq. \eqref{eq:decomp} is evaluated analytically through the Weierstrass elliptic framework, providing a robust and fast baseline. The second term, representing the localized metric deformation near the wormhole throat, is calculated numerically along the identical photon trajectory. Figure~\ref{trajectory} shows the photon trajectories generated using the backward ray-tracing method, with the trajectories containing rays that are deflected around the object as well as rays that approach and may traverse the wormhole throat. 

To ensure numerical stability near the throat and radial turning points ($\mathcal{R}(r) \to 0$), we implemented a segmented integration strategy. The integration path is strictly partitioned at each turning point, and a Simpson-based quadrature with endpoint-nudge regularization is applied to the residual term. This hybrid approach captures the strong-field gravitational effects and localized redshift modifications with high precision, eliminating the systematic errors inherent in simple scaling approximations while preserving the rapid convergence and efficiency of the underlying \textsc{YNOGK} architecture.

\newpage

\section{Model parameter settings}
\label{para_settings}

\begin{table}[ht]
\centering
\renewcommand\arraystretch{1.3}
\caption{\label{bestfit}
Parameters of the \textsc{XSPEC} model \texttt{kwline} for the dotted purple line of Fig. \ref{KerrWH_Feline1}.}
\begin{tabular}{cccc}
\hline\hline
Component & Parameter  & Unit & Value \\
\hline
\texttt{kwline}   & $a_*$                 &      & $0.998$ \\
\texttt{kwline}   & $\theta_{\rm obs}$               & deg  & $75$ \\
\texttt{kwline}   & ${E_\text{line}}$     & \si{\kilo\electronvolt} & $6.40$ \\
\texttt{kwline}   & ${q_\text{in}}$       &      & $4$ \\
\texttt{kwline}   & ${q_\text{out}}$      &      & $4$ \\
\texttt{kwline}   & $r_{\text{br}}$       & $r_g$ & 15 (fixed) \\
\texttt{kwline}   & $r_{\rm in}$          & $r_g$ & $3.32$ \\
\texttt{kwline}   & $r_\textrm{out}$          & $r_g$ & $30$ \\
\texttt{kwline}   & $\lambda$             &      & $0.9$ \\
\texttt{kwline}   & norm                  &      & $1.0$ \\
\hline\hline
\end{tabular}
\label{tab:kwline}
\end{table}

\begin{table}[ht]
\centering
\renewcommand\arraystretch{1.3}
\caption{\label{bestfit}
 Parameter values of the spin angular momentum $a_*$, inclination angle $i$, the emissivity index $q$, break radius $r_\text{br}$, disk inner radius $r_{\rm in}$, disk outer radius $r_{\rm out}$ and throat parameter $\lambda$ of the \texttt{kwconv} component of convolved X-ray spectra, as well as the parameters of the input spectrum component \texttt{xillverCp}.}
\begin{tabular}{cccc}
\hline
Component & Parameter  & Unit & Value \\
\hline
\texttt{kwconv}     & $a_*$            &        & $0.998$ \\
\texttt{kwconv}     & $\theta_{\rm obs}$           & deg    & $75$ \\
\texttt{kwconv}     & ${q_\text{in}}$   &        & $4$ \\
\texttt{kwconv}     & ${q_\text{out}}$  &        & $4$ \\
\texttt{kwconv}     & $r_{br}$          & $r_g$  & 15 (fixed) \\
\texttt{kwconv}     & $r_{\rm in}$             & $r_g$  & $3.32$ \\
\texttt{kwconv}     & $r_{\rm out}$             & $r_g$  & $30$ \\
\texttt{kwconv}     & $\lambda$         &        & $0.9$ \\
\texttt{xillverCp}  & $\Gamma$          &        & $2$ \\
\texttt{xillverCp}  & $A_{\rm Fe}$      &        & $1$ \\
\texttt{xillverCp}  & $\log\xi$         &        & $3.1$ \\
\texttt{xillverCp}  & $kT_{\rm e}$      & \si{\kilo\electronvolt} & $300$ \\
\texttt{xillverCp}  & Dens              &        & $16$ \\
\texttt{xillverCp}  & Incl              & deg    & $75$ \\
\texttt{xillverCp}  & $z$               &        & $0.0$ \\
\texttt{xillverCp}  & norm              &        & $2.6\times10^{-5}$ \\
\hline
\end{tabular}
\label{tab:xspec-params}
\end{table}

\onecolumn

\newpage
\section{Model best-fit results} 
\label{Best-fit}

\begin{table}[ht]
\centering
\caption{{Best-fit parameters obtained by fitting a standard Kerr mock spectrum ($a_* = 0.998$, $i = 75^\circ$) using the customized wormhole reflection model (\texttt{kwconv}) and the modular Kerr model (\texttt{kerrconv}).}}
\label{tab:validation_kerr}
\renewcommand\arraystretch{1.2}
\begin{tabular}{l l c c c}
\hline\hline
Component & Parameter & Unit & \texttt{kwconv} (Wormhole) & \texttt{kerrconv} (Kerr) \\
\hline
Relativistic & $a_*$ & & $0.99985 \pm 0.00345$ & $0.99800 \pm 0.00651$ \\
Convolution & $\theta_{\rm obs}$ & deg & $75.08 \pm 3.97$ & $75.15 \pm 0.21$ \\
& $q_{\rm in}$  & & $3.90 \pm 0.11$ & $3.70 \pm 0.08$ \\
& $q_{\rm out}$  & & 3.90 (=$q_{\rm in}$)  & $10.00 \pm 115.7$ \\
& $r_{\rm br}$ & $r_g$ & $31.89 \pm 3.32$ & $8.10 \pm 10.38$ \\
& $r_{\rm in}$  & $r_g$ & $1.230 \pm 0.066$ & $1.00 \pm 0.08$ \\
& $r_{\rm out}$ & $r_g$ & $27.59 \pm 1.84$ & $399.8 \pm \text{fix}$ \\
& $\lambda$ & & 0.0 & -- \\
\hline
\texttt{xillverCp} & $\Gamma$  & & $1.994 \pm 0.004$ & $1.998 \pm 0.004$ \\
& $A_{\rm Fe}$ &  & $1.001 \pm 0.009$ & $0.982 \pm 0.008$ \\
& $\log \xi$ &  & $3.114 \pm 0.007$ & $3.110 \pm 0.008$ \\
& $kT_e$ & keV & $196.89 \pm 15.50$ & $237.71 \pm 25.79$ \\
& $\log N$  &  & $15.99 \pm 0.03$ & $16.00 \pm 0.03$ \\
& norm & $10^{-5}$ & $2.60 \pm 0.40$ & $0.38 \pm 0.05$ \\
\hline
Statistics & $\chi^2 / \text{d.o.f.}$ & & $1999.65 / 1884 \approx 1.061$ & $1955.68 / 1885 \approx 1.037$ \\
& Null hyp. prob. & & $3.16 \times 10^{-2}$ & $0.126$ \\
\hline\hline
\end{tabular}
\end{table}

\begin{table}[htbp]
\centering
\renewcommand\arraystretch{1.38}
\setlength{\tabcolsep}{4pt}
\caption{{Best-fit parameters for the three simulated spectra. Model 1 (\texttt{kwconv*xillverCp}) and Model 2 (\texttt{kerrconv*xillverCp}) use convolved table models, while Model 3 uses the \texttt{relxillCp} relativistic reflection model. Parameters without uncertainties were frozen during the fit.}}
\label{tab:combined-fitting-results}
\begin{tabular}{lcccc}
\hline\hline
Parameter & Unit & Model 1 (Wormhole) & Model 2 (Kerr BH) & Model 3 (\texttt{relxillCp}) \\ \hline
$n_H$ & $10^{22} \text{ cm}^{-2}$ & - & - & $6.14 \times 10^{-16}$ \\
$a_*$ & & $0.9981 \pm 0.0830$ & $0.9896 \pm 0.0090$ & $0.6876 \pm 0.2457$ \\
$\theta_{\rm obs}$ & deg & $75.03 \pm 3.31$ & $75.50 \pm 0.28$ & $72.01 \pm 1.64$ \\
$q_{\rm in}$  & & $4.03 \pm 0.37$ & $2.96 \pm 0.44$ & $10.00 \pm 94.61$ \\
$q_{\rm out}$  & & $3.13 \pm 7.17$ & $7.63 \pm 10.98$ & $3.11 \pm 0.36$ \\
$r_{\rm br}$  & $r_g$ & $15.28 \pm 41.26$ & $5.76 \pm 3.00$ & $4.79 \pm 4.68$ \\
$r_{\rm in}$  &  $r_g$ & $1.237$ (frozen) & $1.35 \pm 0.07$ & $-1.0$ (frozen) \\
$r_{\rm out}$  & $r_g$ & $29.96 \pm 1.78$ & $381.76$ & $37.98 \pm 12.71$ \\
$\lambda$ & & $0.8991 \pm 0.0006$ & - & - \\
$\Gamma$ & & $2.001 \pm 0.004$ & $2.002 \pm 0.003$ & $2.019 \pm 0.003$ \\
$A_{\rm Fe}$ & & $0.998 \pm 0.009$ & $0.98 \pm 0.04$ & $1.49 \pm 0.03$ \\
$\log\xi$ & & $3.10 \pm 0.06$ & $3.10 \pm 0.05$ & $3.19 \pm 0.01$ \\
$kT_e$ & keV & $282.55 \pm 34.03$ & $279.58 \pm 32.27$ & $284.10 \pm 21.06$ \\
$\log N$ & & $16.00 \pm 0.02$ & $15.99 \pm 0.02$ & $15.00 \pm 0.65$ \\
$refl\_frac$ & & - & - & $10.00 \pm 1.21$ \\
Norm & & $2.70 \times 10^{-5}$ & $3.70 \times 10^{-6}$ & $2.99 \times 10^{-2}$ \\ \hline
$\chi^2 / \text{d.o.f.}$ & & $1920.16 / 1884$ & $1954.18 / 1885$ & $2803.18 / 1916$ \\
Null Hypo. Prob. & & $2.76 \times 10^{-1}$ & $1.30 \times 10^{-1}$ & $1.28 \times 10^{-36}$ \\ \hline\hline
\end{tabular}
\end{table}

\end{appendix}

\end{document}